\documentclass[journal]{IEEEtran}

\ifCLASSINFOpdf
\else
\fi

\usepackage{amsmath,amssymb}
\usepackage[utf8]{inputenc}
\usepackage{graphicx}
\usepackage{subfigure}
\usepackage{booktabs}
\usepackage{witharrows}
\usepackage{rotating}  
\usepackage{multirow}
\usepackage{pifont}
\usepackage{bbding}
\usepackage{adjustbox}
\usepackage{colortbl}
\usepackage{caption}
\usepackage[colorlinks,linkcolor=blue]{hyperref}
\definecolor{Gray}{gray}{0.9}

\def\etal{\emph{et al.}}

\definecolor{Gray}{gray}{0.9}
\def\x{{\mathbf x}}
\def\y{{\mathbf y}}

\newcommand{\ie}{\textit{i.e.}}
\newcommand{\eg}{\textit{e.g.}}
\newcommand{\renjie}[1]{{\color{black}#1}}
\newcommand{\add}[1]{{\color{black}#1}}
\newcommand{\adds}[1]{{\color{black}#1}}

\newcommand{\Fref}[1]{Figure~\ref{#1}}

\newcommand{\renjies}[1]{\textcolor{black}{{#1}}}
\newcommand{\renjiess}[1]{\textcolor{black}{{#1}}}

\begin{document}
%
\title{Enhancing Low-Light Images in Real World via Cross-Image Disentanglement}
%
%
%

\author{Lanqing~Guo, Renjie~Wan,~\IEEEmembership{Member,~IEEE,}
Wenhan~Yang,~\IEEEmembership{Member,~IEEE,}
        Alex~Kot,~\IEEEmembership{Fellow,~IEEE,}
        and~Bihan~Wen,~\IEEEmembership{Member,~IEEE}
\IEEEcompsocitemizethanks{
\IEEEcompsocthanksitem L. Guo, R. Wan, W. Yang, A.C. Kot, and B. Wen are with School of Electrical \& Electronic Engineering, Nanyang Technological University, Singapore 639798.  E-mail: lanqing001@e.ntu.edu.sg, \{rjwan, wenhan.yang, eackot, bihan.wen\}@ntu.edu.sg.
}
}

%
%

\markboth{Journal of \LaTeX\ Class Files,~Vol.~14, No.~8, August~2015}%
{Shell \MakeLowercase{\textit{et al.}}: Bare Demo of IEEEtran.cls for IEEE Journals}
%



\maketitle

\begin{abstract}
\adds{Images captured in the low-light condition suffer from low visibility and various imaging artifacts, \eg, real noise.
Existing supervised algorithms for low-light image enhancement require a large set of pixel-aligned training image pairs, which are hard to prepare in practice. 
Though some recent unsupervised methods can alleviate such data challenges, many real-world artifacts inevitably get falsely amplified in the enhanced results due to the lack of corresponded supervision.
In this paper, instead of using perfectly aligned images for training, we creatively employ the misaligned real-world images as the guidance, which are considerably easier to collect.
Specifically, we propose a Cross-Image Disentanglement Network (CIDN) with weakly supervised learning, to separately extract cross-image brightness and image-specific content features from low/normal-light images.} Based on that, CIDN can simultaneously correct the brightness and suppress image artifacts in the feature domain, which largely increases the robustness of the pixel shifts.
Furthermore, we collect a new low-light image enhancement dataset consisting of misaligned training images with real-world corruptions.
Experimental results show that our model achieves state-of-the-art performances on both the newly proposed dataset and other popular low-light datasets.
\end{abstract}


%
\IEEEpeerreviewmaketitle

\section{Introduction}
\label{sec:Introduction}
\begin{figure}[!t]
	\begin{center}
		\begin{tabular}{c@{ }c}
			\includegraphics[width=.47\linewidth]{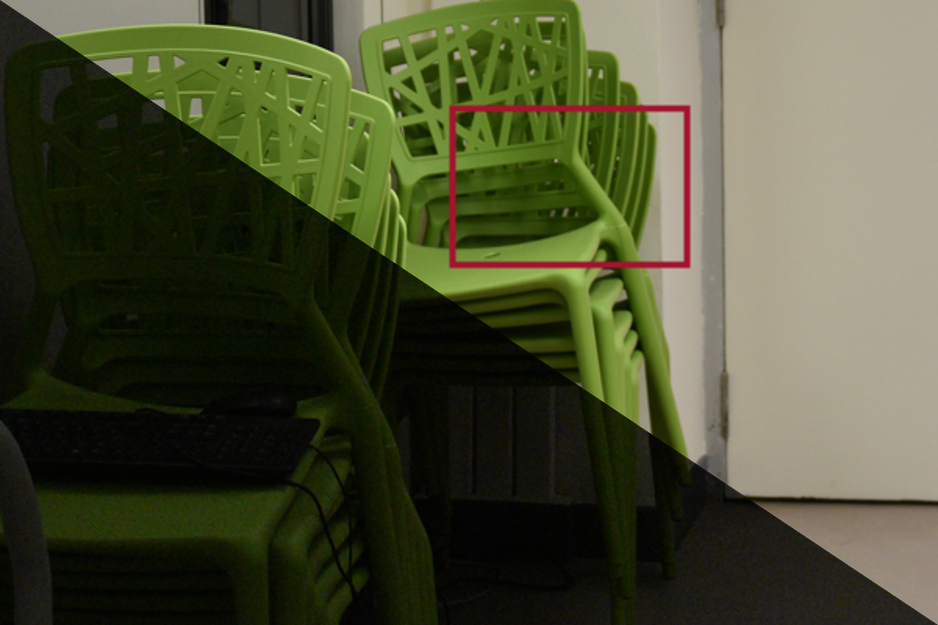}~&
			\includegraphics[width=.47\linewidth]{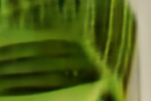}\\
			(a) Input+GT~& (b) KinD (Fully Supervised)\\
			\includegraphics[width=.47\linewidth]{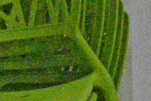}~&
			\includegraphics[width=.47\linewidth]{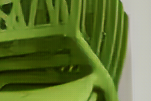}\\
			(c) Zero-DCE (Unsupervised)~& (d) Ours\\
		\end{tabular}
	\end{center}
\caption{
Examples of the enhanced low-light images from the proposed NM2L dataset:
(a) Input low-light image (bottom-left) and its corresponding normal-light groudn truth (top-right), and the enhanced results using (b) KinD~\protect\cite{zhang2019kindling} (fully supervised method), (c) Zero-DCE~\protect\cite{guo2020zero} without paired supervision (unsupervised method), and (d) our proposed CIDN.
}
\label{fig:intro} 
\end{figure}
Images captured under low-light conditions are ubiquitous in real-world scenes, \eg, night-time surveillance and autonomous-driving. They suffer from poor visibility, such as low contrast, low intensity, and high ISO noise, limiting both human perception and the performances of subsequent vision tasks (\eg, object detection~\cite{Exdark} and Re-ID~\cite{xu2020black}). 
Thus, it is essential to restore low-light images for better visibility and usability in practice.

Recent low-light enhancement methods rely on learning deep neural networks from a large-scale training corpus, which achieved promising performance.
The fully supervised methods~\cite{lore2017llnet,Chen2018Retinex,zhang2019kindling} learn to remove the related composite degradation based on the well-aligned paired low/normal-light images and elaborately designed models.
Some of them~\cite{Chen2018Retinex,zhang2019kindling} follow the classic Retinex theory~\cite{land1977retinex} to learn pixel-wise reflectance and illumination decomposition in a data-driven way.
\renjies{Most learning-based methods are sensitive to the misalignment between training pairs as the constraints of the applied pixel-wise reflectance and illumination decomposition or the pixel-aligned loss functions.} 
However, low-light images with paired and pixel-aligned ground truth are scarce in practice.  
\adds{On one hand, the time-varying sensor setup (\eg, unstable camera devices) makes it difficult to obtain the perfectly aligned ground truth~\cite{yamamoto2012subjective}, which is a common challenge in practice amongst various computational photography tasks, \eg, reflectance removal~\cite{zhang2019zoom}, medical image translation~\cite{mccord1992misalignment,kong2021breaking}, and remote sensing~\cite{faiza2012review}.}
On the other hand, \adds{the images captured in the real world} might include scene or object movements. As a result, even slight pixel shifts between the training pairs lead to severe performance degradation using fully supervised methods~\cite{Chen2018Retinex,zhang2019kindling}, as shown in~\Fref{fig:intro}(b). 
Though one can synthesize image pairs that are perfectly aligned for training, their distribution inevitably deviates from real-world images due to the domain gap, resulting in artifacts when \adds{applying the trained models to real testing images}.

\begin{figure*}[ht] 
\centering 
\includegraphics[width=1.\linewidth]{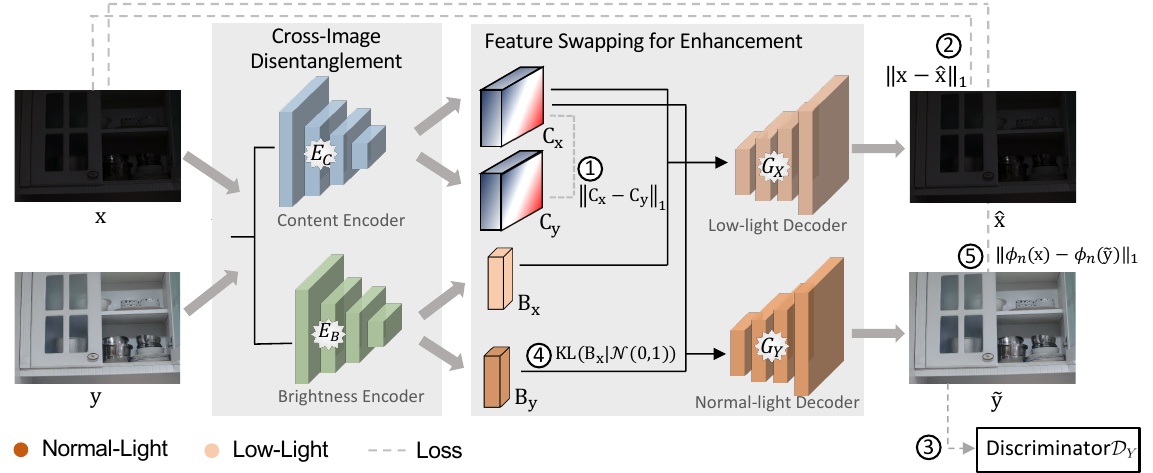} 
\caption{Architecture of the proposed Cross-Image Disentanglement Network (CIDN). The overall architecture is divided into two components: \textit{cross-image disentanglement} (left) consists of the content encoder $E_C$ and the brightness encoder $E_B$ extracting cross-image content and brightness features from input low/normal-light images, respectively;
\textit{feature swapping for enhancement} (right) consists of the low-light decoder $G_X$ reconstructing the low-light image (right top) and the normal-light decoder $G_Y$ (right bottom) re-combining the swapped brightness from normal-light image and content features from low-light image for enhancement.
\ding{172}\ding{173}\ding{174}\ding{175}\ding{176} represent the different losses utilized to optimize the whole network: \ding{172} content consistency loss, \ding{173} reconstruction loss, \ding{174} multi-scale adversarial loss, \ding{175} KL loss, and \ding{176} perceptual loss, respectively.} 
\label{fig:pipeline_2} 
\end{figure*}

\adds{To alleviate such data challenge, recent methods employ the unsupervised training strategies for low-light image enhancement, \eg, Zero-DCE~\cite{guo2020zero} and EnlightenGAN~\cite{jiang2021enlightengan}.} However, the noise and artifacts in the low-light images may inevitably be amplified during the illumination or brightness correction process. Without corresponded guidance/supervision, these methods can hardly suppress the amplified noise, as shown in Figure~\ref{fig:intro}(c). Thus, how to correct the image brightness and suppress artifacts based on reliable and easily accessible guidance poses the unique challenges for low-light image enhancement problems in the real-world.

\adds{In this work, we introduce a weakly-supervised setting for low-light image enhancement, \ie, the training set contains pairs of low/normal-light images that are misaligned for low-light image enhancement.} By relaxing the requirement for perfectly aligned images in supervised learning, it becomes easier to collect large-scale training images from the real world under our setup.
To handle the misaligned training pairs, we propose a novel Cross-Image Disentanglement Network (CIDN) to encode low/normal-light images to low-dimensional feature spaces. 
In such feature spaces, the encoded misalignment-mitigated representations from reference images can be used as the training guidance.
As shown in~\Fref{fig:pipeline_2}, CIDN first introduces a cross-image disentanglement \renjies{with} the brightness and content encoders \renjies{to} extract two agnostic features, \ie, the \textit{brightness feature} containing cross-image structure-independent brightness information and the \textit{content feature} containing complete structure-aware information.
The feature disentanglement enables CIDN to simultaneously enhance the image brightness and suppresses artifacts in the feature domain by \textit{brightness} \textit{feature swapping} and \textit{content consistency} \textit{refinement}, respectively.
\add{With the merits of cross-image disentanglement, low-light images can be enhanced with any \textbf{arbitrary} normal-light guidance image at the inference stage.}
Furthermore, we collect a new dataset of misaligned real-world image pairs to verify the effectiveness of our method. Our major contributions can be concluded as follows:


\begin{itemize}
     \item \add{A dataset for low-light image enhancement with real noise and misalignment between the low/normal-light image pairs.}
     
    \item \renjie{A cross-image disentanglement framework to enhance low-light images \adds{using weakly-supervised learning}, \ie, training with misaligned low/normal-light image pairs.}
    \item \renjie{A feature swapping strategy with content consistency constraint to correct image brightness and suppresses artifacts simultaneously.}
    
    \item \add{A feasible solution to adjust the illumination with a arbitrary normal-light guidance image in inference stage.}

\end{itemize}

The rest of the paper is organized as follows: Section II introduces related low-light image enhancement methods, unpaired image-to-image translation methods, and image denoising methods. In Section III, the proposed dataset is introduced and compared with existing low-light datasets. In Section IV, the proposed approach is introduced and analyzed, and the algorithm is described. Experimental results are shown in Section V and concluding remarks are given in Section VI.

\section{Related Work}
\subsection{Low-Light Image Enhancement}
Many researchers have explored low-light image enhancement task,
which aims to enlarge the visibility for subsequent classification, detection, and recognition. 
The earliest low-light enhancement methods, such as
Histogram equalization (HE)~\cite{abdullah2007dynamic,lee2013contrast}, spread out the most frequent intensity values to achieve uniformly contrast improvement.
Such global illumination adjustment without local adaptation easily leads to undesirable over/under-exposure and intensive noise.
Later on,
Retinex theory~\cite{land1977retinex}, assuming the image can be decomposed into reflectance and illumination, has been widely used in traditional illumination-based methods~\cite{jobson1997multiscale,fu2016fusion,guo2016lime}.
For instance,
LIME~\cite{guo2016lime} applies a structure-aware smoothing regularization to estimate the illumination map.
Fu \etal~\cite{fu2016fusion} adjusts the illumination through fusing multiple derivations of the initially estimated illumination map.
These methods can remove slight noise in images while handling heavy noise and artifacts with only hand-craft priors.



Recently, deep learning based methods apply high-quality normal-light ground truth as guidance to learn how to improve low-light image~\cite{Chen2018Retinex,lore2017llnet,gharbi2017deep,zhang2019kindling}. 
LL-Net~\cite{lore2017llnet} makes the first attempt by proposing a stacked auto-encoder to simultaneously conduct denoising and enhancement using synthesized low/normal-light image pairs. However, the distribution of synthetic data inevitably deviates from real-world images due to the domain gap, leading to severe performance degradation when transferring to real-world cases.
\adds{Later on, Wei~\etal~\cite{Chen2018Retinex} collects a real-world dataset with low/normal-light image pairs, based on which proposes Retinex-Net for pixel-wise decompose images into illumination and reflectance maps in a data-driven way, which additionally employs BM3D~\cite{dabov2006image} as the postprocessing denoiser to deal with noise in the reflectance map. Following that, Zhang~\etal~\cite{zhang2019kindling} proposes KinD to jointly train the illumination enhancement and reflectance denoising modules. However, those methods decomposes illumination and reflectance maps in the spatial domain, which is a highly ill-posed problem. Some structural information would be wrongly decomposed into an illumination map, which might result in unnatural output. Different from them, in this paper, we propose a novel disentanglement in deep feature domain, which largely increases the robustness to noise and pixel-shift between training pairs.}


More recently, inspired by unsupervised image translation methods, \cite{jiang2021enlightengan} proposed to directly enlighten low-light images without any paired training data. Particularly, \cite{guo2020zero} only exploited internal properties of the image to enhance the intensity.
\cite{zheng_2020_ECCV} targeted to adverse weather condition and focused more on the accuracy of the subsequent high-level tasks rather than the quality of generated images. 
However, these methods are not robust to low-light images with obvious real noise or extremely dark cases due to lacking a reliable guidance. 

\subsection{Unpaired Image-to-Image Translation}
Image-to-image translation aims to translate images from the source domain to the corresponding target domain.
Many computer vision tasks can be posed as this problem, \eg, Long~\etal~\cite{long2015fully} proposes a fully convolutional network (FCN) for image-to-segmentation translation. SRGAN~\cite{ledig2017photo} maps low-resiolution images to high resolution images. EnlightenGAN~\cite{jiang2021enlightengan} translates low-light images to normal-light one via global and local discriminators.
Since it is usually impractical to collect aligned training data, unpaired learning based algorithms have been widely adopted.
With the merit of adversarial training, Dumoulin~\etal~\cite{dumoulin2016adversarially} and Donahue~\etal~\cite{donahue2016adversarial} propose algorithms to jointly learn mappings between
latent space and data bidirectionally.
The well-known CycleGAN~\cite{zhu2017unpaired} first applies the cycle-consistency loss to attempt to train with unpaired data, which learns an inverse mapping from the output domain back to the input and checks if the input can be reconstructed. 
UNIT~\cite{liu2017unsupervised} makes a shared-latent space assumption based on coupled GANs. 
As following, to improve the diversity of output, models such as MUNIT~\cite{huang2018multimodal}, DRIT~\cite{lee2018diverse} are proposed to embed images onto domain-invariant content space and domain-specific attribute space via disentanglement. 
Choi~\etal~[10] further proposed a
StarGAN that can perform image-to-image translations for
multiple domains using only a single model. Similarly, Liu~\etal~\cite{liu2018unified} proposed a UFDN that learns domain-invariant
representation from multiple domains and can perform continuous cross-domain image translation and manipulation.
More recently, some other works also utilized contrastive learning for more controllable image-to-image translation. Park~\etal~\cite{park2020contrastive} encourages content preservation by maximizing the mutual information between input and output with contrastive learning. DivCo~\cite{liu2021divco} deals with mode collapse issue in conditional generative adversarial networks via latent-augmented contrastive loss.

\subsection{Image Denoising}
Image denoising is a typical ill-posed problem, with numerous techniques proposed over past decades.
It is dedicated to recovering high-quality images from their noisy measurements, which also improves robustness in various high-level vision tasks.
Classic methods take advantage of image priors, such as sparsity~\cite{elad2006image}, low rank~\cite{gu2014weighted}, and non-local self-similarity~\cite{dabov2007color,liu2018non}. 
The representative works, such as BM3D~\cite{dabov2006image} applying effective filtering in 3D transform domain by combining sliding-window transform processing with block matching. 
WNNM~\cite{gu2014weighted} incorporating low-rank matrix approximations using the weighted nuclear norm.
Lately, deep learning based denoisers exhibits superiority to learn image models from training dataset with an end-to-end approach~\cite{zhang2017beyond,zhang2018ffdnet,liu2018non}. 
For instance, Zhang~\etal~\cite{zhang2017beyond} achieves very competitive denoising performance with residual learning.
Followed that, \cite{zhang2018ffdnet} introduces a noise level map to control the trade-off between noise reduction and detail preservation.
To exploit the non-local property of the image features in deep convolutional neural network, Plotz~\etal~\cite{Ploetz:2018:NNN}
presents an N3Net by employing the k-nearest neighbor
matching in the denoising network.
Noise is prevalent in low-light images due to the low signal-to-noise ratio (SNR). 
The enhanced normal-light image corrupted by spatially variant noise, \ie, different regions of an enhanced normal-light image are corrupted by different levels of noise.
Simple pre/post-processing using existing denoising modules is sub-optimal due to complicated noise distribution and dual loss of structural details. 
Different from regarding denoising as a pre/post-processing, our CIDN introduces to simultaneously correct the brightness and suppress the noise via disentanglement. 

\section{Dataset Preparation}

\begin{figure}
\centering
\includegraphics[width=1.\linewidth]{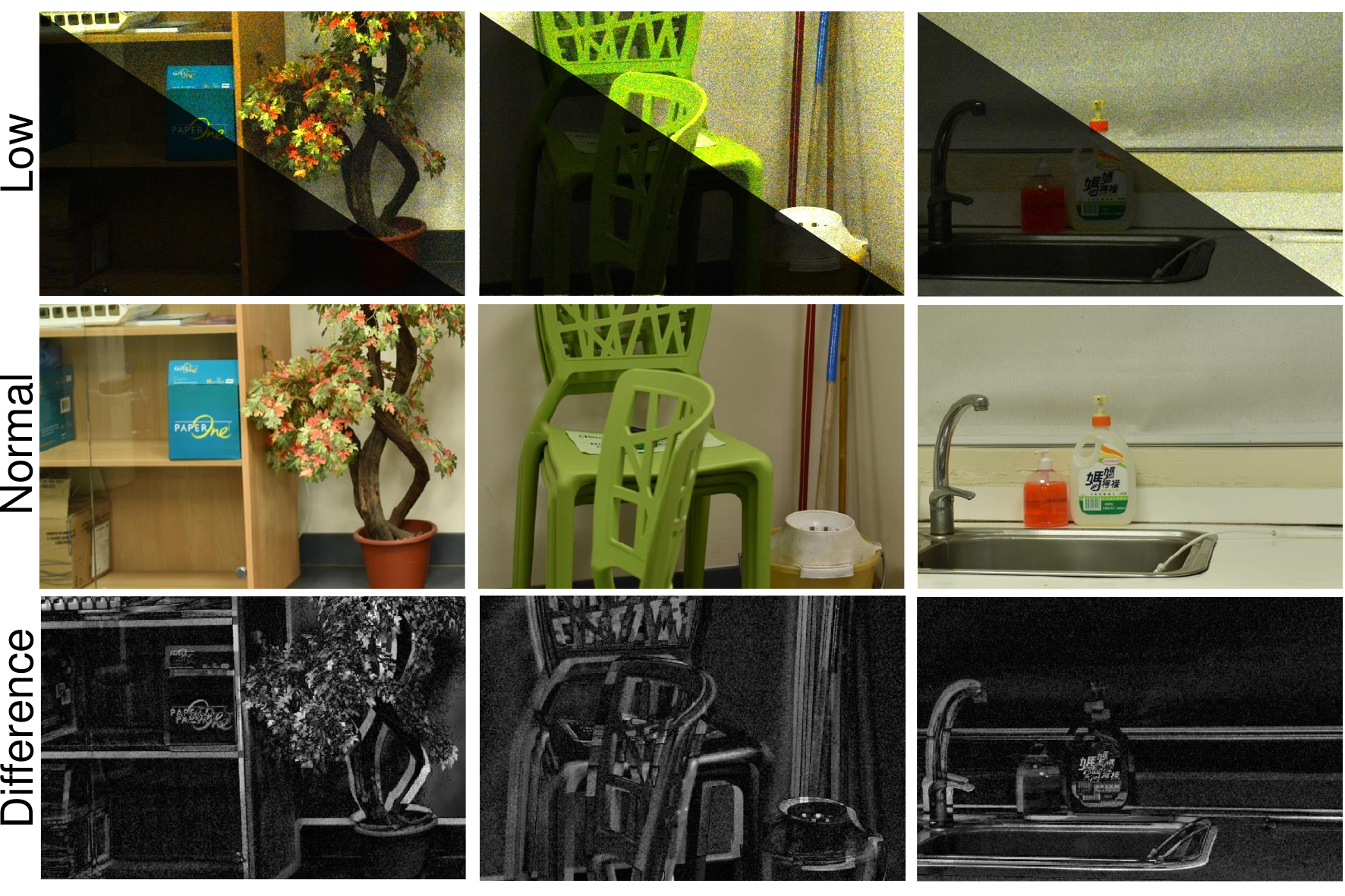} 
\caption{Examples of NM2L dataset: First row shows the low-light images with the corresponding exposure adjusted results; Second row shows the corresponding normal-light images; Third row demonstrates the difference between image pairs (the overlapping is colored black and the difference is colored white). Note that the low-light $\times 10$ version is on the top-right corner of the first row for better visibility.} 
\label{fig:nm2l_example} 
\end{figure}

Existing low-light datasets contain unpaired images, or pairs of synthetic images, or limited real-world images.
%
Though a time-consuming three-step algorithm~\cite{anaya2018renoir} can be applied to eliminate the misalignment~\cite{Chen2018Retinex,anaya2018renoir}, it largely reduces both the quantity and diversity of the collected images in the dataset which is not scalable. 
To complement such limitation, we propose a Noisy Misaligned Low-Light (NM2L) dataset by facilitating the use of misaligned training pairs of real images. 
Table~\ref{tab:datasets_comp} compares the proposed NM2L dataset to the existing popular datasets for low-light image enhancement.
\renjie{In contrast to datasets like LOL~\cite{Chen2018Retinex}, our training pairs are captured by first taking a normal-light image with normal lightness, and then turning down the environmental lightness or camera sensitivity to capture the corresponding low-light image.}
Without the use of a tripod, remote control, or any lossy post-processing step, the misaligned image pairs are much easier to collect, and thus more scalable.  
Figure~\ref{fig:nm2l_example} shows some examples of our NM2L dataset, with the misalignment being visualized in the third row (white regions).
Furthermore, different from existing low-light datasets like LOL~\cite{Chen2018Retinex} and LIME~\cite{guo2016lime}
taken by good photographic skills with low noise, low-light images from NM2L have obvious real noise measured at low-light environments. 
Thus, NM2L can be applied to (1) train enlightening models for more realistic scenarios, and (2) benchmark various enlightening algorithms under a real-world setup.

\begin{table}[!t]
\centering
\footnotesize
\setlength{\tabcolsep}{0.5em}
  \caption{Comparisons between existing public low-light datasets, including LOL~\cite{Chen2018Retinex}, {SID\protect\footnotemark}~\cite{chen2018learning}, LIME~\cite{guo2016lime}, RENOIR~\cite{anaya2018renoir}, and NM2L (our proposed dataset).}
\adjustbox{width=0.9\linewidth}{
    \begin{tabular}{c|ccccc}
        \toprule
        \textbf{Datasets} & LOL & SID & LIME & RENOIR & \cellcolor{Gray}NM2L \\
    \midrule
    Image Pairs & \Checkmark & \Checkmark  & & & \cellcolor{Gray}\Checkmark \\ 
    Obvious Real Noise &   &   &  & \Checkmark &\cellcolor{Gray}\Checkmark \\
    Misalignment & & & & & \cellcolor{Gray}\Checkmark\\
        \bottomrule
    \end{tabular}
}
\label{tab:datasets_comp}
\end{table}
\footnotetext{SID dataset is a raw image dataset, containing short-exposure and its corresponding long-exposure image pairs. The authors fixed the suitable (low) ISO when taking the short-exposure and long-exposure pairs. Thus, short-exposure images (low-light images) from SID can be regarded as almost noise-free image.}

\adds{In total, our training dataset contains $1000$ low/normal-light image pairs with $1600\times 2400$ resolutions at diverse scenes,  which significantly increases the data richness and complexity for training. We resize and crop the original images into  smaller patches with $400 \times 600$ resolutions for training and testing.}
To better evaluate the performances of our proposed algorithm, we also follow the standard strategy to capture an evaluation dataset with $50$ image pairs.
\adds{
We also notice that the ground truth normal-light images in testing set of the LOL dataset have inconsistent brightness, since the normal-light images are captured under different environmental lights, \eg, different times of the day. It might result in incorrect measurements during inference stage since the image quality metrics, \eg, PSNR and SSIM, are very sensitive to brightness.
To this end, to achieve a correct and fair benchmark, we try to keep the brightness of ground truth images consistent by adjusting the camera parameters.
}
Moreover, different from the training dataset with misaligned pixels, images of the evaluation dataset are better aligned for a more accurate evaluation.

\section{Proposed Method}

We first introduce our proposed Cross-Image Disentanglement Network (CIDN). Figure~\ref{fig:pipeline_2} shows the architecture of CIDN, which can be divided into two major components, \ie, cross-image disentanglement to extract brightness and content features from low/normal-light image pairs, and feature swapping for image enhancement in the latent space.
After that, we formulate the loss functions for optimizing CIDN. 
We also give some examples and discussion on how our CIDN works during inference stage.

\subsection{Cross-Image Disentanglement Network}
The goal of the proposed CIDN is to learn a mapping from low-light domain $\mathcal{X}$ to normal-light domain $\mathcal{Y}$, which 
can be separated into the cross-image disentanglement (left) and feature swapping for enhancement (right) as shown in Figure~\ref{fig:pipeline_2}. 
The cross-image disentanglement consists of a content encoder $E_C$ mapping images into content space $(E_C : \mathcal{X}, \mathcal{Y} \rightarrow \mathcal{C})$ and a brightness encoder $E_B$ mapping images into brightness space $(E_B : \mathcal{X}, \mathcal{Y} \rightarrow \mathcal{B})$.
The feature swapping for enhancement consists of a low-light decoder $G_X$ generating low-light images using low-light brightness features and content features $(G_X: \{\mathcal{C},\mathcal{B}\} \rightarrow \mathcal{X})$, a normal-light decoder $G_Y$ generating a normal-light image using normal-light brightness features and content features $(G_Y: \{\mathcal{C},\mathcal{B}\} \rightarrow \mathcal{Y})$,
 and two domain discriminator sets $\mathcal{D}_X$ and $\mathcal{D}_Y$ with multi-scale discriminators to hierarchically detect whether the generated images come from the generator or from the real data distribution.

\vspace{1mm}
\noindent\textbf{Cross-image disentanglement.} \renjie{Given a low-light image $\x$ and its corresponding misaligned normal-light image $\mathbf{y}$, we encode them into the corresponding content space $\mathcal{C}$ and brightness space $\mathcal{B}$ as:}
\begin{align}
    \nonumber & \{\mathbf{C_x}, \mathbf{C_y}\} = \{E_C(\mathbf{x}), E_C(\mathbf{y})\} \quad \mathbf{C_x}, \mathbf{C_y} \in \mathcal{C}\\
    & \{\mathbf{B_x}, \mathbf{B_y}\} = \{E_B(\mathbf{x}), E_B(\mathbf{y})\} \quad \;\mathbf{B_x}, \mathbf{B_y} \in \mathcal{B}\;,
\end{align}
where $\{\mathbf{C_x}, \mathbf{C_y}\}$ and $\{\mathbf{B_x}, \mathbf{B_y}\}$ denote the content and brightness features for $\x$ and $\y$, respectively. 
The encoded content features contain all of the structure-aware information, while the encoded brightness features is structure-independent.
The feature embedding of both low/normal-light images share the same brightness and content encoders, as they have the common brightness and content feature spaces.


\vspace{1mm}
\noindent\textbf{\renjie{Feature swapping for enhancement.}}  Since the low-light image $\x$ and normal-light image $\mathbf{y}$ involve pixel shifts, the optimization based on pixel-wise constraints may lead to additional artifacts. We eliminate the influence from pixel misalignment by utilizing the disentangled feature information based on the translation-invariant property of convolutional neural network~\cite{jaderberg2015spatial}. As shown in \Fref{fig:pipeline_2}, we correct the image brightness based on the feature swapping strategy, \ie, combining the brightness features $\mathbf{B_y}$ from normal-light image and the content features $\mathbf{C_x}$ from a low-light image \renjie{as follows:} 
\begin{equation}
    \Tilde{\mathbf{y}} = G_Y (E_C(\mathbf{x}) , E_B(\mathbf{y}))\;,
\end{equation}
\renjie{where $\Tilde{\mathbf{y}}$ is the enhanced result for $\mathbf{x}$, and $G_Y$ denotes the normal-light decoder used to reconstruct the enhanced image.} 

\begin{figure*}[!ht]
\centering
 \vspace{-1mm}
\includegraphics[width=1.\linewidth]{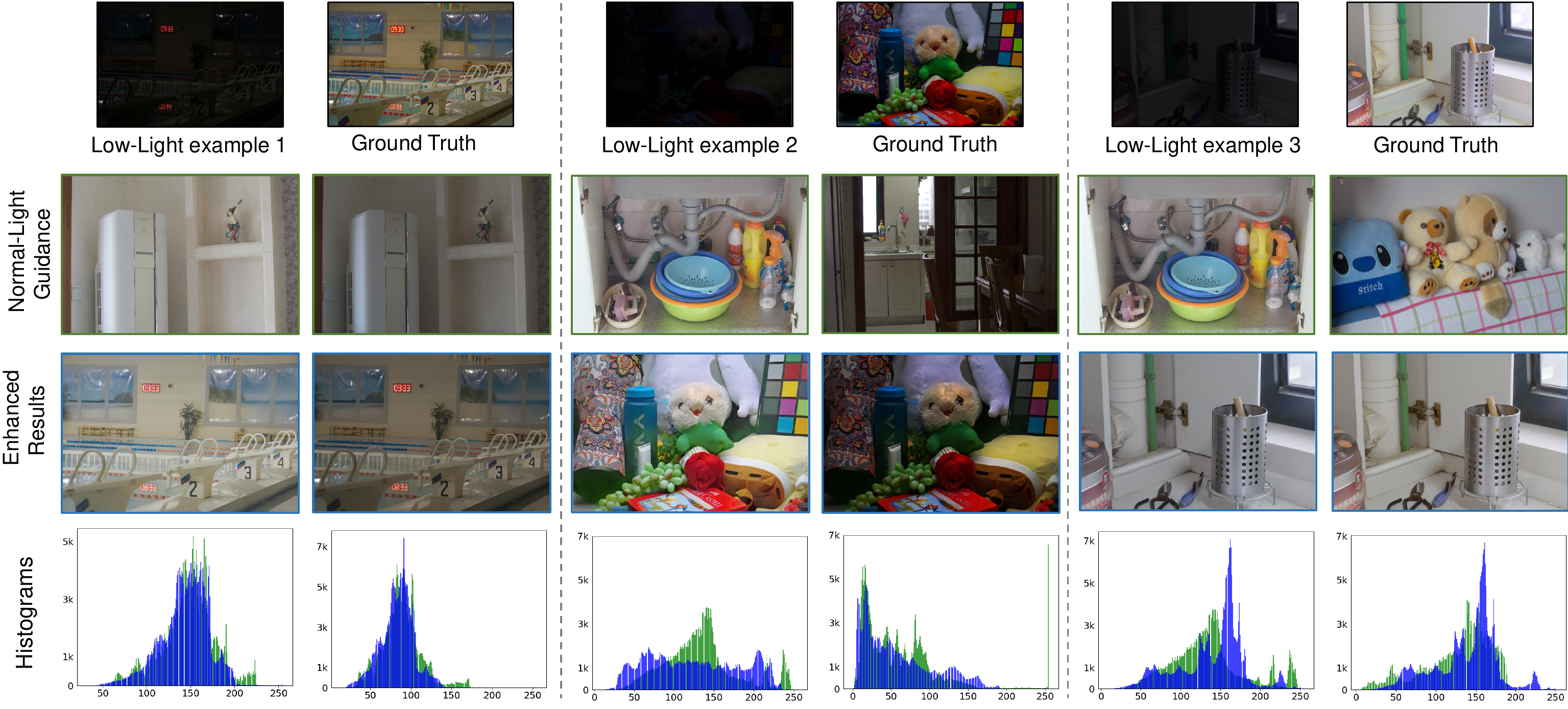} 
%
\caption{Three different examples for low-light image restoration under the guidance of specific normal-light image during the inference stage, (1) example 1: two normal-light guidance are taken in the same scene but with different lightness, (2) example 2: two normal-light guidance are taken in the different scene with different lightness, (3) example 3: two normal-light guidance are taken in the different scene with similar lightness.
The first row is input low-light image and its corresponding normal-light ground truth. The second row is different normal-light guidance. The third row is enhanced results by corresponding guidance. The forth row is the histogram of normal-light guidance (\textbf{{\color{green}green}}) and enhanced result (\textbf{{\color{blue}blue}}).
Please \textbf{ZOOM} \textbf{IN} to see the details.} 
\label{fig:visual} 
\end{figure*}

\subsection{Losses for Training}
\vspace{1mm}
\noindent\textbf{Content consistency loss.}
\renjie{Though the feature swapping strategy for enhancement can correct the image brightness, the artifacts may not necessarily be suppressed, \eg, $\mathbf{C_x}$ may still contain noise. 
To this end, we propose a content consistency loss by employing the noise-free content features $\mathbf{C_y}$ as the reference.} Since the content is invariant to different light conditions~\cite{land1977retinex}, the content features $\{\mathbf{C_x}, \mathbf{C_y}\}$ \renjie{extracted from images with similar scenes should} contain similar information. \renjie{Though the pixel shift between our training pairs may influence this consistency, the translation invariance of a convolutional neural network can alleviate this issue.} Thus, to attain the potential feature consistency, we add a content consistency loss using $\ell_1$ loss on $\{\mathbf{C_x}, \mathbf{C_y}\}$ as 
\begin{equation}
\mathcal{L}_{con}=\left\|\mathbf{C_x}-\mathbf{C_y}\right\|_1\;.
\end{equation}

\vspace{1mm}
\noindent\textbf{Reconstruction loss.}
To align the generated output with the corresponding input image pairs, we enforce the extracted brightness features $\mathbf{B_x}$ and content features $\mathbf{C_x}$ from the low-light image can reconstruct the original low-light image after the decoder, and similarly reconstruct the normal-light image by $\mathbf{B_y}$ and $\mathbf{C_y}$:
\begin{align}
    \hat{\mathbf{x}} = G_X(\mathbf{C_x}, \mathbf{B_x}) \quad \hat{\mathbf{y}} = G_Y(\mathbf{C_y}, \mathbf{B_y})\;.
\end{align}
We adopt the classical $\ell_1$ loss as the image reconstruction loss to reload the input image pair using all encoded features, with the goal of reconstructing $\mathbf{x}$, as follows:
\begin{equation}
\mathcal{L}^{\mathcal{X}}_{rec} =  \left\|\hat{\mathbf{x}}-\mathbf{x}\right\|_1,
\end{equation}
 where $\hat{\mathbf{x}}$ denotes the reconstructed low-light image. The image reconstruction loss provides an external incentive for the content encoder to extract crucial structural information and ensures the integrity of the information contained in the extracted brightness and content features.
 Similarly define $\mathcal{L}_{rec}^{\mathcal{Y}}$ to enforce the reconstruction of $\mathbf{y}$.

\vspace{1mm}
\noindent\textbf{Multi-scale adversarial loss.}
In order to enforce all brightness variant information embedded to brightness features,
we impose multi-scale adversarial loss to globally and locally judge the generated quality on both domains. Take the low-light domain as an example,
\begin{equation}
\mathcal{L}_{adv}^{\mathcal{X}} =\min _{E_{B},E_{C}, G_X} \max _{D_{X}^k \in \mathcal{D}_X} \sum_{k=1,2,3} \mathcal{L}_{\mathrm{GAN}}^{\mathcal{X}}\left(E_{B}, E_{C}, G_X,D_X^k\right)\;\renjie{,}
\end{equation}
where $\mathcal{D}_X$ is the set of discriminators, $D_{X}^k (k=1,2,3)$ denotes three discriminators for different image scales and the largest one is always global discriminator, and $\mathcal{L}_{\mathrm{GAN}}^{\mathcal{X}}$ is in the form of vanilla adversarial loss~\cite{goodfellow2014generative}.



\vspace{1mm}
\noindent\textbf{Other losses.}
We also employ a KL loss to assume Gaussian prior for the distribution of latent brightness codes, 
so that the encoded brightness features can be further enforced to be independent of structure information, which is defined as:
\begin{equation}
\mathcal{L}^{\mathcal{X}}_{\mathrm{KL}} = \mathrm{KL}(\mathbf{B_x}\parallel \mathcal{N}(0,1))\;\renjie{,}
\end{equation}
where $\mathrm{KL}(\cdot)$ denotes the KL-divergence that penalizes deviation of the latent distribution from the Gaussian prior.
Thus, images can be massively generated by their internal content and sampling brightness features.


Furthermore, to preserve the perceptual details after encoders and decoders, we employ a perceptual loss as
\begin{equation}
\mathcal{L}_{per}^{\mathcal{X}}= \sum_{n=1}\frac{1}{K_n} \|\phi_{n}\left(\mathbf{x}\right)-\phi_{n}\left(\Tilde{\mathbf{y}}\right)\|_1\;,
\end{equation}
which makes the generated image $\Tilde{\mathbf{y}}$  perceptually similar to $\mathbf{x}$~\cite{jiang2021enlightengan}, and similarly apply $\mathcal{L}_{per}^{\mathcal{Y}}$ to mitigate the semantic difference between the generated dark image $\Tilde{\mathbf{x}}$ and input normal-light image $\mathbf{y}$.
Here $\phi_n$ denotes the $n$-th layer feature map of the pretrained VGG-16 model on ImageNet and $K_n$ indicates the number of activations in that layer.

By combining the above losses, the hybrid objective function $\mathcal{L}$ used to train our model is
\begin{align}
\nonumber\mathcal{L} = \; & \mathcal{L}^{\mathcal{X}}_{rec} + \mathcal{L}^{\mathcal{Y}}_{rec} + w_1\mathcal{L}_{con} + w_2(\mathcal{L}^{\mathcal{X}}_{\mathrm{KL}} + \mathcal{L}^{\mathcal{Y}}_{\mathrm{KL}}) \\
+\;& w_3(\mathcal{L}_{per}^{\mathcal{X}} 
 + \mathcal{L}_{per}^{\mathcal{Y}}) 
 + w_4 (\mathcal{L}_{adv}^{\mathcal{X}} + \mathcal{L}_{adv}^{\mathcal{Y}})\;,
\end{align}
where $w_1$, $w_2$, $w_3$, and $w_4$ are the weighting coefficients to balance the influence of each term.


\subsection{Inference Stage}
\adds{Due to the lack of paired reference images with similar contents at the inference stage, the brightness features can come from two ways: (1) a selected guidance image with the desired normal brightness, or (2) zero vector brightness feature to obtain the information of reflectance map. Because of the cross-image property of our disentangling framework, the brightness information can be accurately disentangled from the guidance even with distinct contents.}

\adds{
\textbf{First}, 
since the cross-image disentangle framework provides a feasible solution to adjust the illumination, the brightness features can come from any \textit{arbitrary} guidance image at the inference stage.
}
Thus, the proposed CIDN provides a user-friendly way to arbitrarily adjust the brightness level by inputting guidance with the expected brightness.
Figure~\ref{fig:visual} shows the examples of enhanced images using different guidance images: the final enhanced results are with similar brightness levels to the guidance and its structure and content information are consistent with the low-light images. Besides, the histograms of the enhanced images (blue) are well aligned with those of the guidance images (green), which demonstrate the effectiveness of the cross-image disentangling framework.
\adds{
\textbf{Second}, the brightness feature can also be fixed as zero vector since the content feature should include complete visible structural information, which is consistent under different lightness.  Inspired by some traditional retinex-based works~\cite{fu2016fusion,guo2016lime} that regarded the reflectance map as the optimal normal-light result, we reconstruct the normal-light image with only the extracted content information and drop the brightness features.
}

\section{Experimental Results}

\subsection{Implementation Details}
\adds{Our CIDN model is implemented in PyTorch\footnote{The reproducible implementations and the collected dataset will be available soon.}, which is conducted on one RTX A5000 GPU.
 The training process lasts for 600 epochs with around 8 hours. We set the mini batch size as 8.}
\adds{The detailed architectures of brightness encoder $E_B$ and content encoder $E_C$ are provided in the supplementary document.}
The extracted brightness features are fixed as $1\times 1\times 8$ that are far smaller than content features, since content contains richer texture and edges information.
We adopt the ADAM optimizer with an initial learning rate of $0.0001$.
 We fix the hyper-parameters $w_2=0.001$, $w_3 = 0.1$ and $w_4 = 1$, and $w_1$ is empirically set based on noise level for obtaining good trade-offs between noise removal and detail preservation (\renjie{\eg, $w_1=1$ for dataset with noise-free low-light images and $w_1=0.2$ for our NM2L with noisy low-light images.}). The network parameters are initialized randomly. During training, we randomly crop patches of resolution $256\times 256$ from the scaled images of resolution $400\times 600$.

\subsection{Performance Evaluation}
We evaluate our method on the proposed Noisy Misaligned Low-Light (NM2L) dataset. \renjie{Furthermore, we also test our CIDN over the LOL dataset~\cite{Chen2018Retinex}, in which the training image pairs are well-aligned with mild noise\footnote{
\adds{We also evaluate our method on some popular non-reference datasets, \ie, DICM~\cite{lee2012contrast}, MEF~\cite{ma2015perceptual}, LIME~\cite{guo2016lime}, ExDark~\cite{Exdark}, DarkFace~\cite{poor_visibility_benchmark}, and LLIV-Phone~\cite{LoLi}. Results of non-reference datasets can be found in supplementary document.}}.}
\adds{We choose the non-learning based method LIME~\cite{guo2016lime}, unsupervised methods EnlightenGAN~\cite{jiang2021enlightengan}, LR3M~\cite{ren2020lr3m}, and Zero-DCE~\cite{guo2020zero}, fully supervised methods Retinex-Net~\cite{Chen2018Retinex},RUAS~\cite{liu2021ruas}, CSDNet~\cite{ma2021learning},  Zhao~\etal~\cite{Zhao_2021_ICCV}, KinD~\cite{zhang2019kindling}, Lv~\etal~\cite{lv2021attention}, and MIRNet~\cite{Zamir2020MIRNet}}\footnote{
\adds{We only report the result of MIRNet on aligned LOL testing set since it does not provide the training code for image enhancement.}}, and semi-supervised method DRBN~\cite{Yang_2020_CVPR} as the competitors. 
We use Peak Signal-to-Noise Ratio (PSNR) and Structural SIMilarity (SSIM) for quantitative measurement. 
In NM2L dataset, though there may be mild misalignment in the evaluation image pairs, this misalignment exists across all methods and thus the comparisons are fair~\cite{zhang2019zoom}.
We also adopt the recently proposed LPIPS~\cite{zhang2018perceptual} as the enhancement quality metric, which measures the perceptual image similarity between the enhanced and ground truth images using a pretrained deep network. The lower LPIPS value indicates better performance.

\begin{figure*}[!t]
\centering
\includegraphics[width=1.\linewidth]{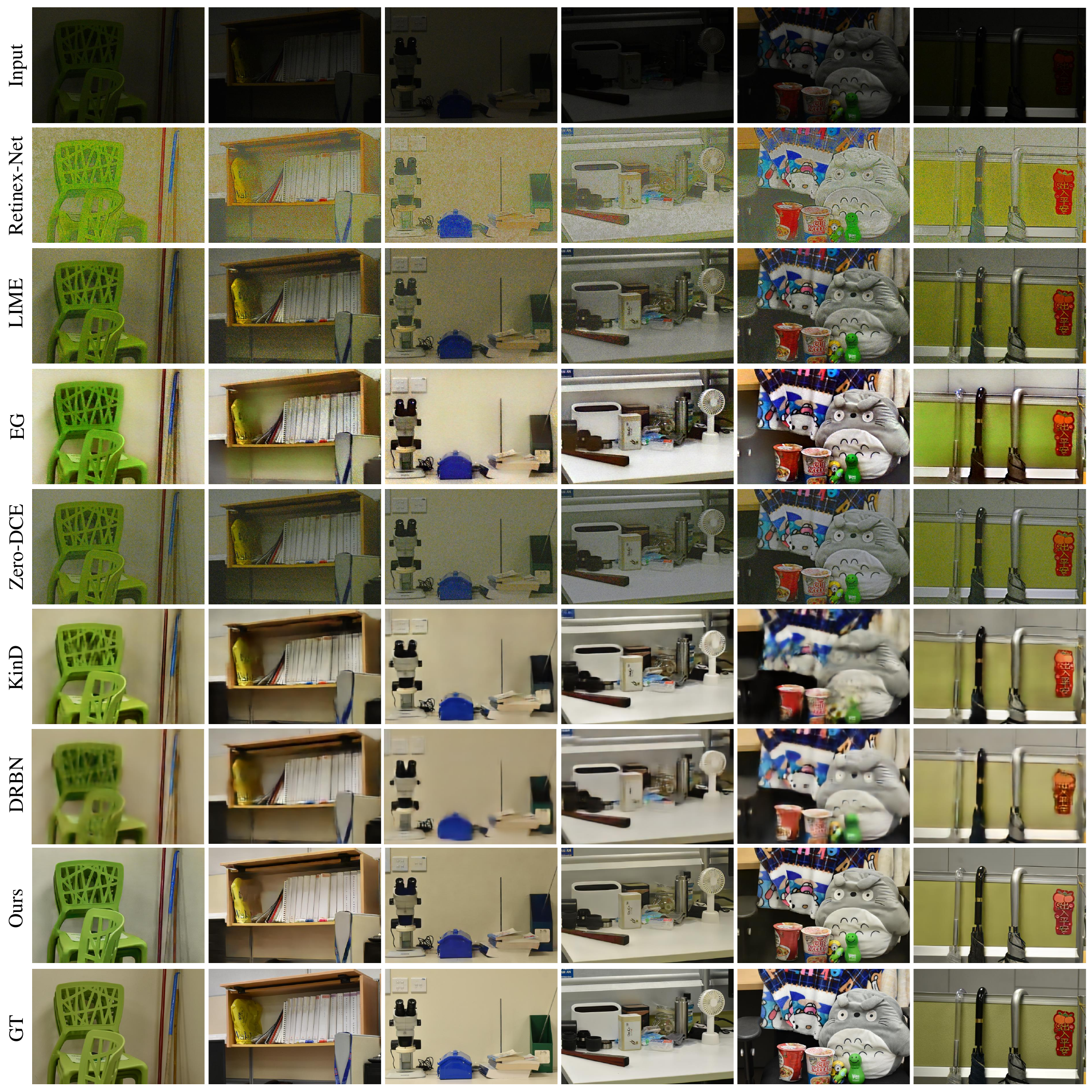} 
\caption{Examples of enhancement results on the evaluation dataset. From top to bottom, the input low-light image, the estimated results of Retinex-Net~\protect\cite{Chen2018Retinex}, LIME~\protect\cite{guo2016lime}, EnlightenGAN (EG)~\protect\cite{jiang2021enlightengan},  Zero-DCE~\protect\cite{guo2020zero}, KinD~\protect\cite{zhang2019kindling}, DRBN~\cite{Yang_2020_CVPR} and our method, and corresponding ground truth. Please \textbf{ZOOM IN} to see the details.} 
\label{fig:nm2l_res} 
\end{figure*}

\begin{figure*}[!t]
	\begin{center}
		\begin{tabular}{c@{ }c@{ }c@{ }c@{ }c@{ }c}
			\includegraphics[width=.19\linewidth]{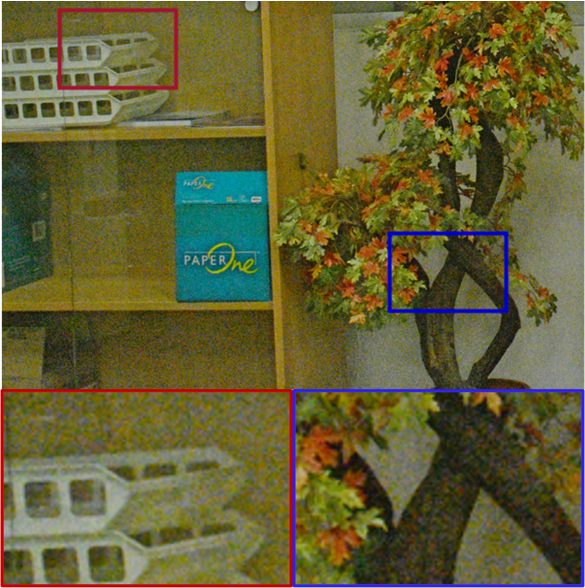}~&
			\includegraphics[width=.19\linewidth]{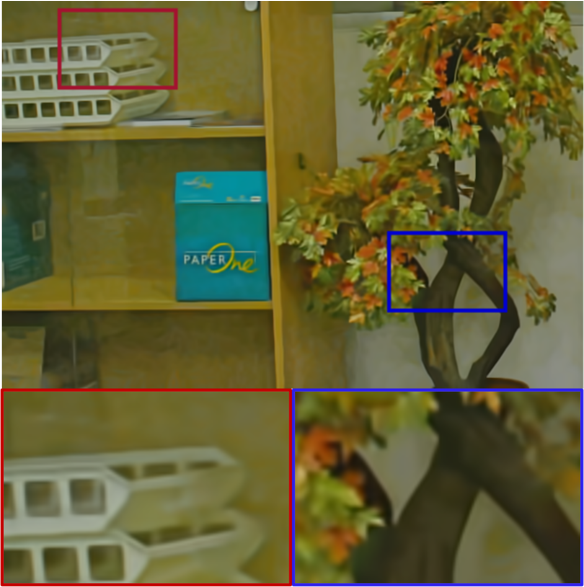}~&
			\includegraphics[width=.19\linewidth]{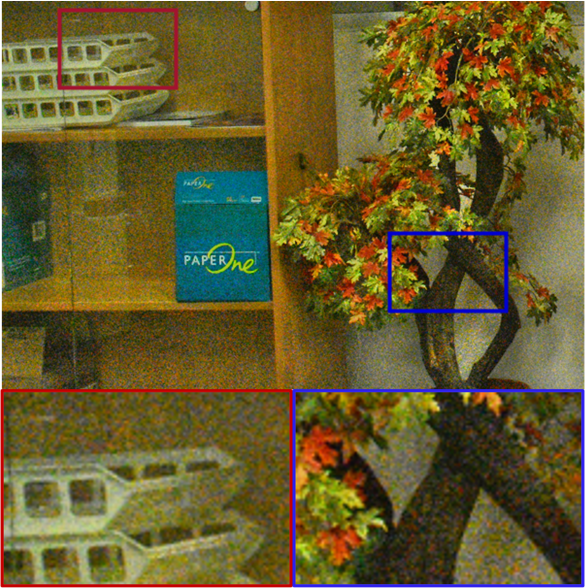}~&
			\includegraphics[width=.19\linewidth]{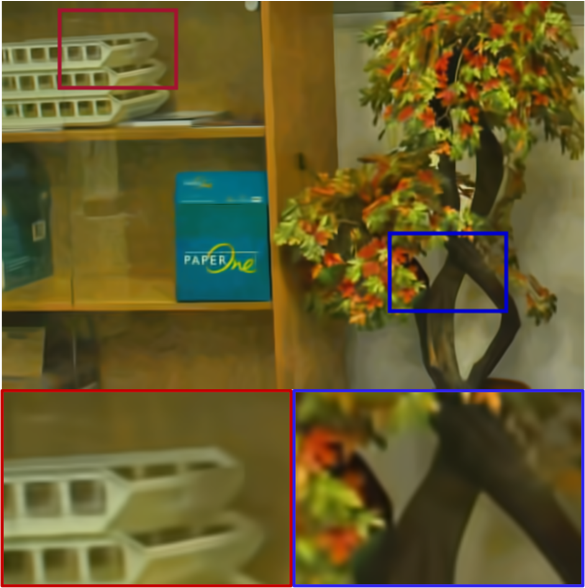}~&
			\includegraphics[width=.19\linewidth]{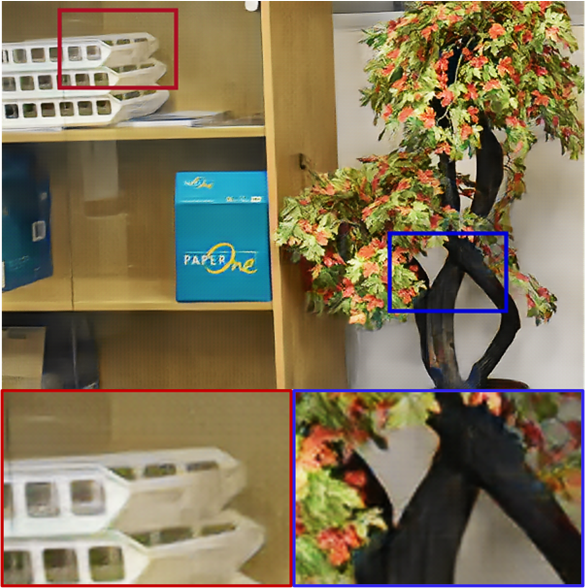}\\
				(a) LIME~& (b) LIME + CBM3D ~& (c) Zero-DCE~&  (d)  Zero-DCE+CBM3D~& (e) Ours \\
		\end{tabular}
	\end{center}
\caption{Visual results obtained by low-light enhancement methods: (a) LIME and (c) Zero-DCE, results using low-light enhancement and CBM3D as post-processing denoising methods (b) and (d), and result of our method (e). Below are the zoom-in regions for better visibility.
}
\label{fig:noisy} 
\end{figure*}

\begin{table}[!t]
\centering
\footnotesize
\setlength{\tabcolsep}{0.5em}
\caption{Quantitative results on our NM2L dataset. +CBM3D denotes employing an additional CBM3D~\protect\cite{dabov2007color} denoiser as post-processing to enhanced results.}
\adjustbox{width=0.8\linewidth}{
    \begin{tabular}{l|ccc}
        \toprule
         Method & PSNR $\uparrow$  & SSIM $\uparrow$ & LPIPS $\downarrow$\\\midrule
         LIME~\cite{guo2016lime} & 12.95  & 0.42  & 0.65 \\ 
        LIME+CBM3D & 13.30  & 0.71 & 0.34\\
        Retinex-Net~\cite{Chen2018Retinex} &14.59  & 0.31 & 0.76 \\
        EnlightenGAN~\cite{jiang2021enlightengan} & 17.02 & 0.66 &0.37 \\
        Zero-DCE~\cite{guo2020zero} &13.58  & 0.51 &0.57  \\
        Zero-DCE+CBM3D &13.87 & 0.72 & \textcolor{blue}{0.33} \\
          KinD~\cite{zhang2019kindling} & 16.20 &  0.56 & 0.38 \\
          DRBN~\cite{Yang_2020_CVPR} & \textcolor{blue}{16.71} & \textcolor{blue}{0.59} & 0.46 \\
         \rowcolor{Gray} Ours & \textcolor{red}{\textbf{21.58}} & \textcolor{red}{\textbf{0.81}} & \textcolor{red}{\textbf{0.16}}\\
        \bottomrule
    \end{tabular}
}
\label{tab:nm2l_res}
\end{table}

\vspace{1mm}
\noindent\textbf{Misaligned and real noisy dataset.}
We first evaluate our method on the proposed Noisy Misaligned Low-Light (NM2L) dataset.
Except for LIME which is a classic method, thus does not require any training data, other competing methods are all trained on our NM2L dataset. 
From the quantitative results shown in Table~\ref{tab:nm2l_res}, our method outperforms all competing methods by large margins. \renjie{The higher SSIM and PSNR values indicate less reconstruction error between the enhanced output and ground truth.} 
Besides, the lower LPIPS result demonstrates that the proposed method provides better visually pleasing results. 
Figure~\ref{fig:nm2l_res} shows some examples of the enhanced images using the proposed CIDN, comparing to those obtained by competing methods.
It is clear that CIDN generates higher-quality images, which not only corrects the brightness but also effectively suppresses the noise and artifacts. In contrast, the classic method, \ie, LIME~\cite{guo2016lime}, and unsupervised method, \ie, Zero-DCE~\cite{guo2020zero}, usually amplify image noise falsely during enhancement.
Unlike the supervised methods, \ie, Retinex-Net~\cite{Chen2018Retinex} and KinD~\cite{zhang2019kindling}, and semi-supervised method, \ie, DRBN~\cite{Yang_2020_CVPR}, which overfits the pixel misalignment during training, the proposed CIDN effectively alleviated the ghosting or over-smoothing artifacts in the enhancement results.
Note that it is challenging to remove the amplified noise or artifacts during post-processing, due to their highly complicated and non-uniform distribution. 
To illustrate the challenge, we apply a popular denoiser CBM3D~\cite{dabov2007color} to the enhanced results with amplified noise using the competing methods, LIME, and Zero-DCE, named LIME+CBM3D and Zero-DCE+CBM3D. The PSNRs of the denoised results are also reported in Table~\ref{tab:nm2l_res}, denoted as LIME+CBM3D and Zero-DEC+CBM3D, respectively, which only provide minor improvement.
Besides, Figure~\ref{fig:noisy} shows that post-processing, though suppresses noise with a certain level, usually generates over-smoothed image details that degrades the visual quality. 
Dual loss makes the high-frequency information reduced, and the noise is still not well suppressed.
More visual examples and implementation details are included in the supplementary document.



\begin{figure*}[!t]
	\begin{center}
		\begin{tabular}{c@{ }c@{ }c@{ }c}
			\includegraphics[width=.23\textwidth]{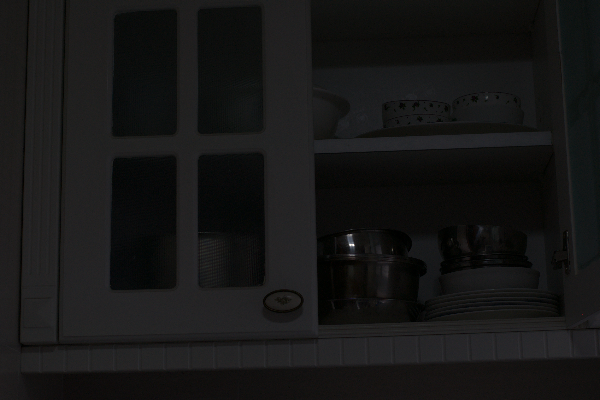}~&
			\includegraphics[width=.23\textwidth]{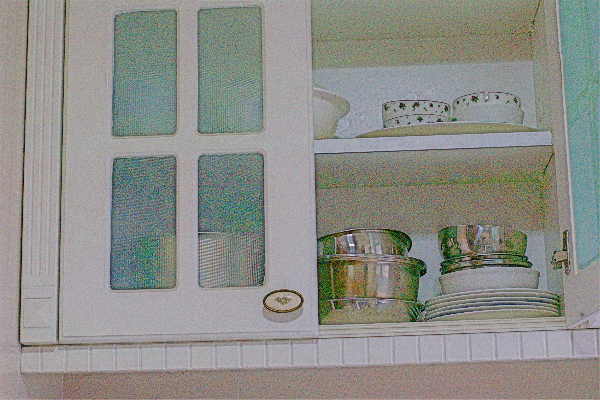}~&
			\includegraphics[width=.23\textwidth]{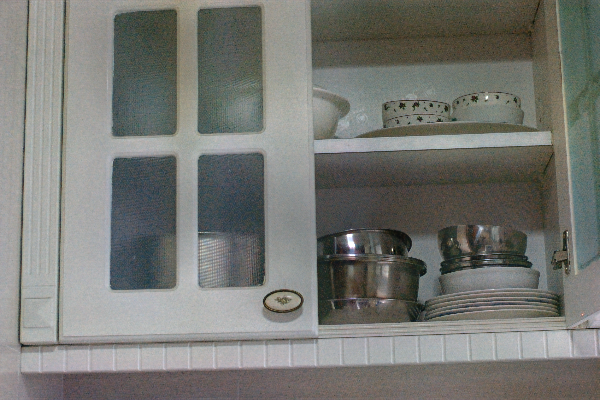}~&
			\includegraphics[width=.23\textwidth]{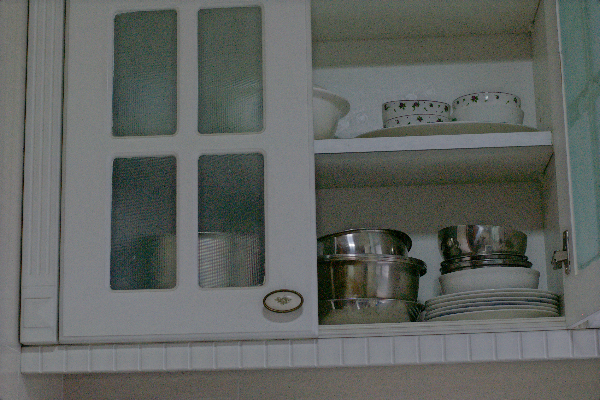}\\
			 Input~&  RetinexNet~\cite{Chen2018Retinex}~&  EnlightenGAN~\cite{jiang2021enlightengan}~&  Zero-DCE~\cite{guo2020zero}\\
			\includegraphics[width=.23\textwidth]{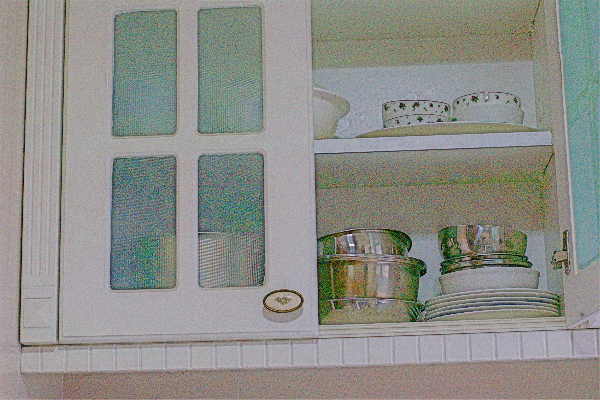}~&
			\includegraphics[width=.23\textwidth]{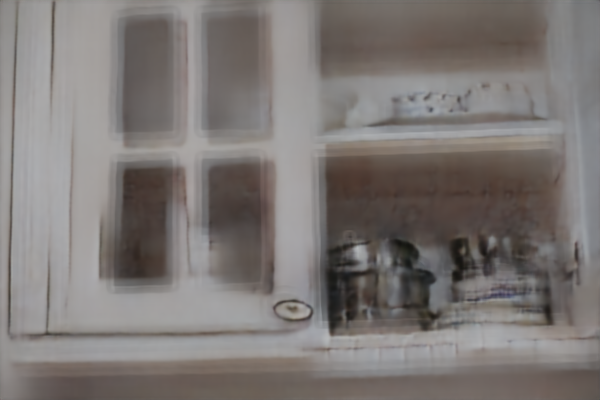}~&
			\includegraphics[width=.23\textwidth]{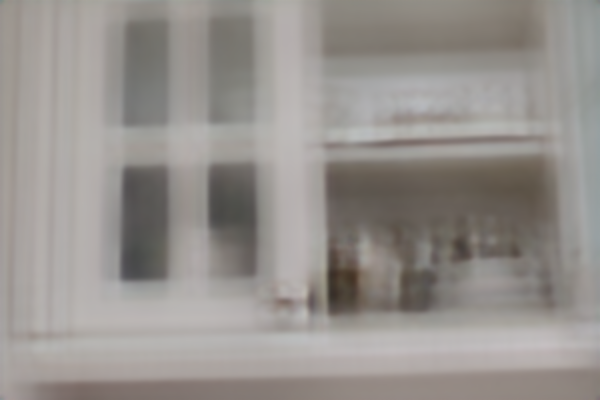}~&
			\includegraphics[width=.23\textwidth]{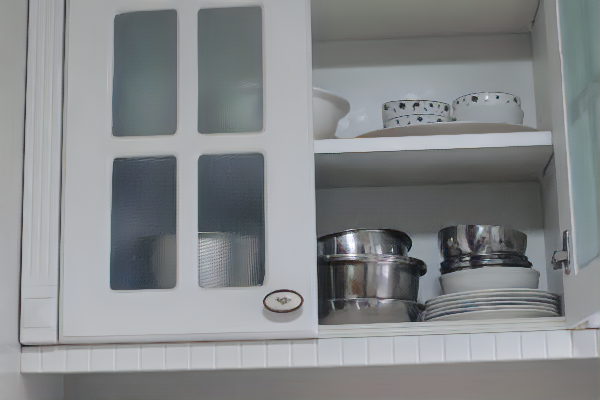}\\
						 Retinex-Net~\cite{Chen2018Retinex}-M~&  KinD~\cite{zhang2019kindling}-M~&  DRBN~\cite{Yang_2020_CVPR}-M~&  \textbf{Ours-M}\\
			\includegraphics[width=.23\textwidth]{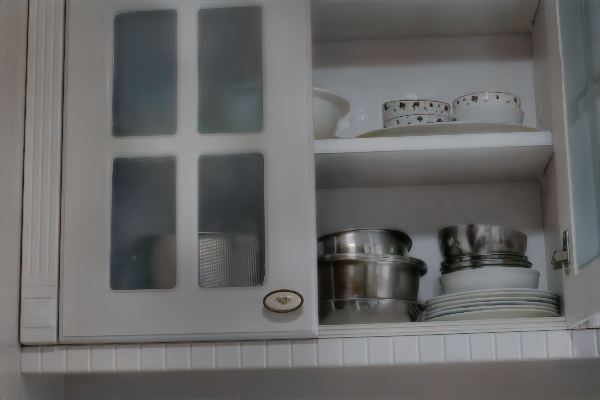}~&
			\includegraphics[width=.23\textwidth]{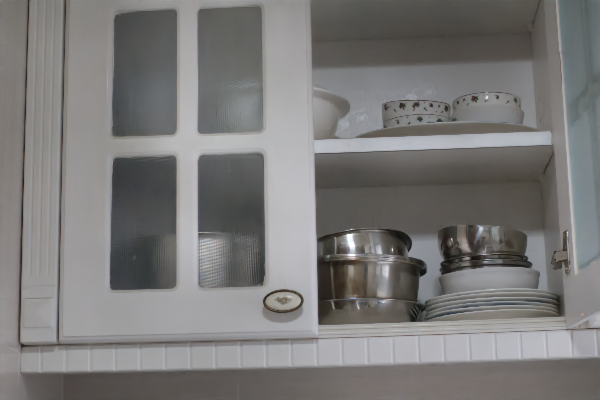}~&
			\includegraphics[width=.23\textwidth]{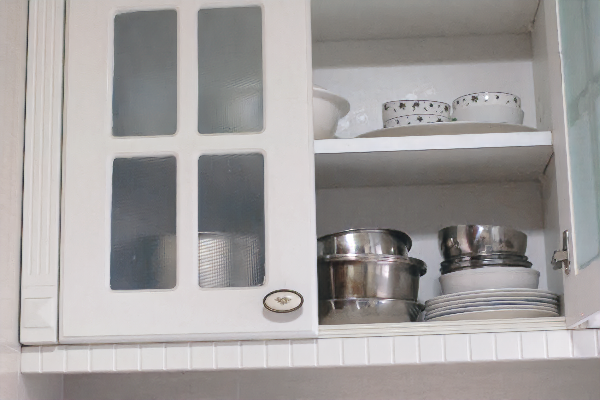}~&
			\includegraphics[width=.23\textwidth]{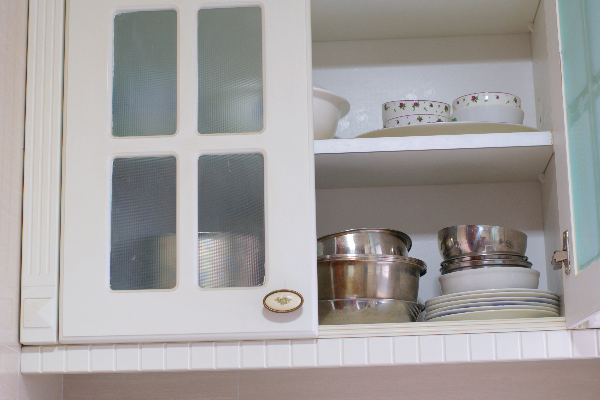}\\
			 KinD~\cite{zhang2019kindling}~&  DRBN~\cite{Yang_2020_CVPR}~& \textbf{Ours}~&  GT\\
						\includegraphics[width=.23\textwidth]{img/LOL_1/input.png}~&
			\includegraphics[width=.23\textwidth]{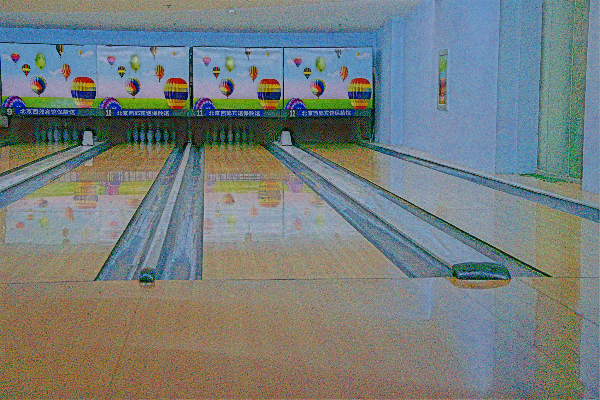}~&
			\includegraphics[width=.23\textwidth]{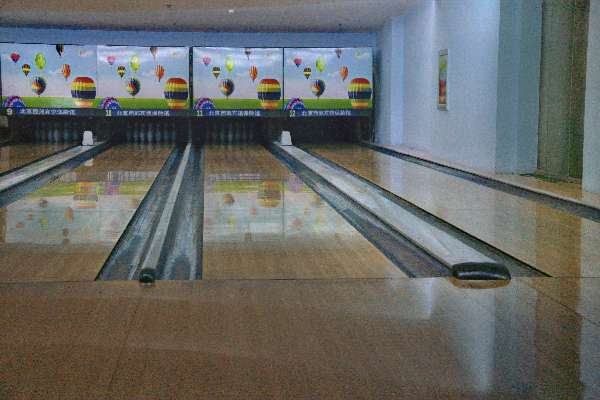}~&
			\includegraphics[width=.23\textwidth]{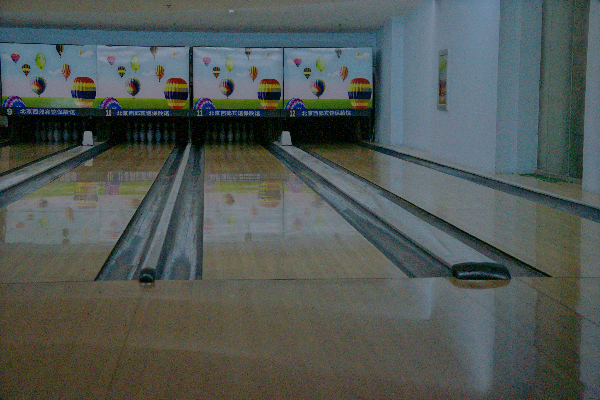}\\
			 Input~&  RetinexNet~\cite{Chen2018Retinex}~&  EnlightenGAN~\cite{jiang2021enlightengan}~&  Zero-DCE~\cite{guo2020zero}\\
			\includegraphics[width=.23\textwidth]{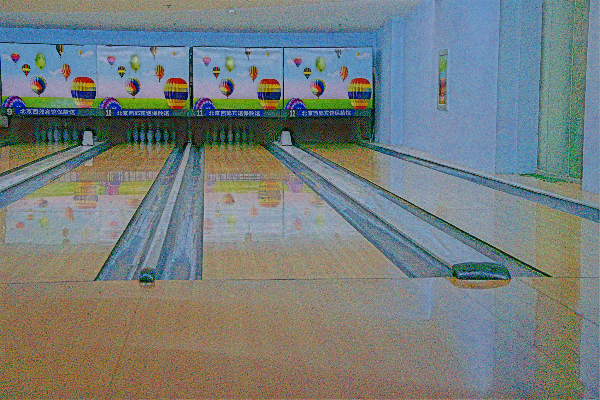}~&
			\includegraphics[width=.23\textwidth]{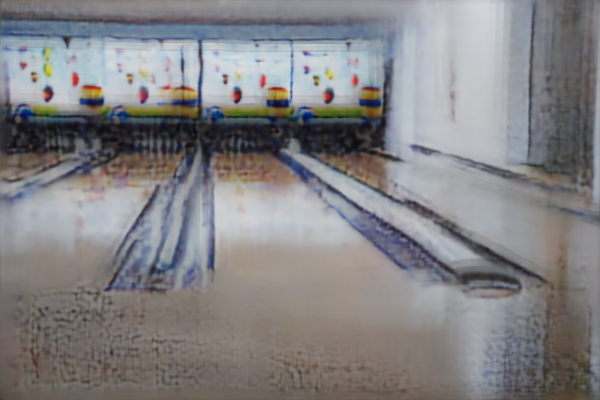}~&
			\includegraphics[width=.23\textwidth]{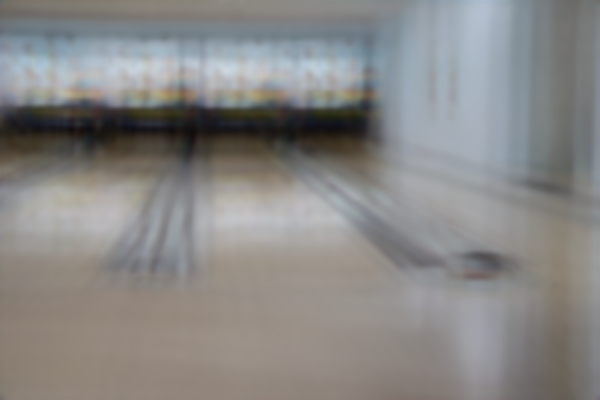}~&
			\includegraphics[width=.23\textwidth]{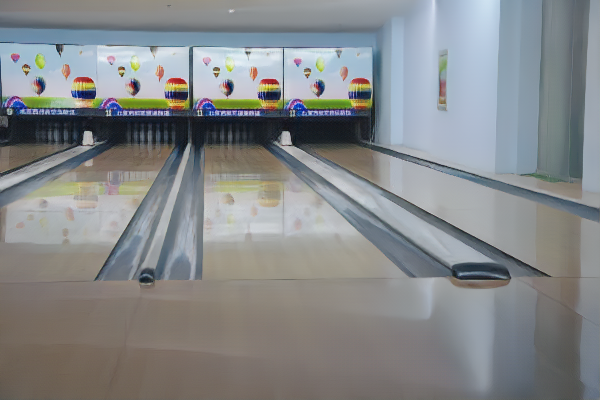}\\
						 Retinex-Net~\cite{Chen2018Retinex}-M~&  KinD~\cite{zhang2019kindling}-M~&  DRBN~\cite{Yang_2020_CVPR}-M~&  \textbf{Ours-M}\\
			\includegraphics[width=.23\textwidth]{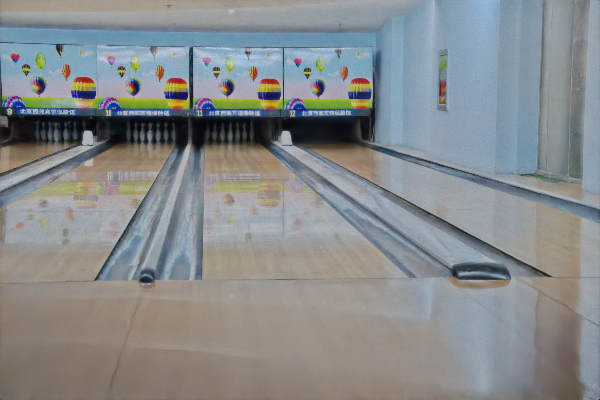}~&
			\includegraphics[width=.23\textwidth]{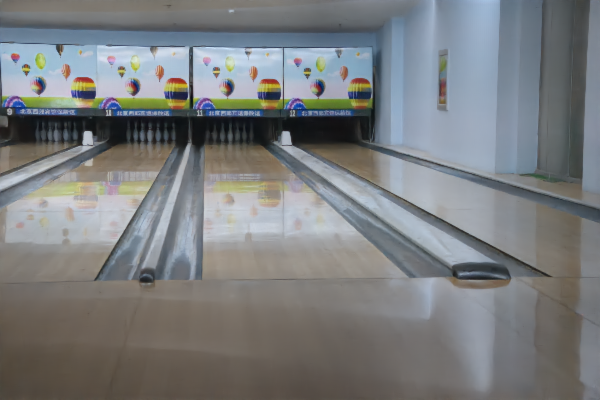}~&
			\includegraphics[width=.23\textwidth]{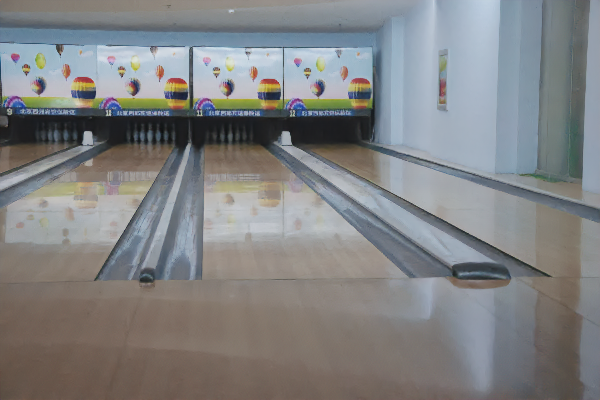}~&
			\includegraphics[width=.23\textwidth]{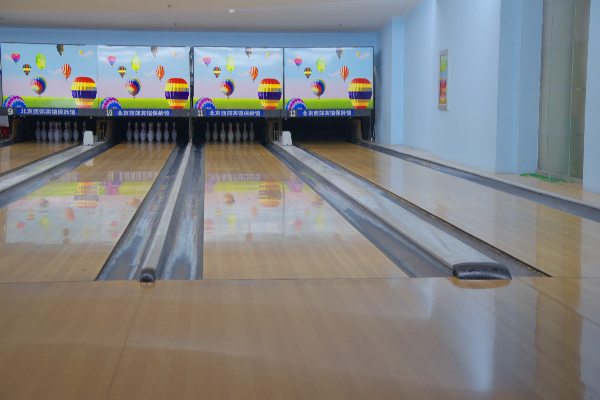}\\
			 KinD~\cite{zhang2019kindling}~&  DRBN~\cite{Yang_2020_CVPR}~&  \textbf{Ours}~& GT\\
		\end{tabular}
	\end{center}
	\caption{Examples of enhancement results on the LOL~\cite{Chen2018Retinex} evaluation dataset. -M denotes these methods are training with misaligned image pairs, while others training with perfectly aligned image pairs. Please \textbf{ZOOM IN} to see the details.}
	\label{fig:cleanlol}
\end{figure*}

\begin{figure*}[t]
	\begin{center}
		\begin{tabular}{c@{ }c@{ }c@{ }c@{ }c@{ }c}
			\includegraphics[width=.23\textwidth]{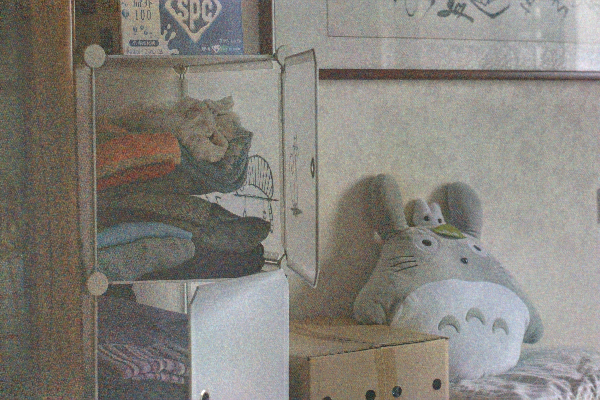}~&
			\includegraphics[width=.23\textwidth]{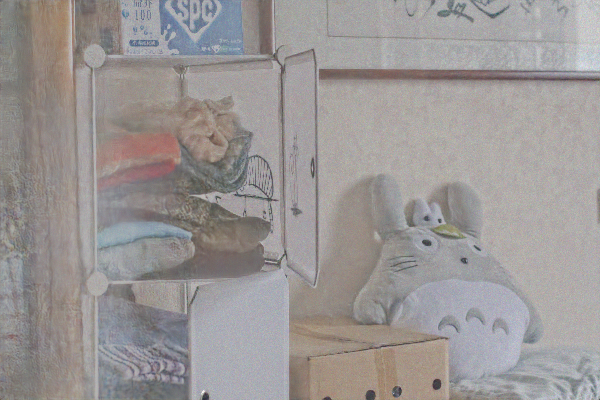}~&\includegraphics[width=.23\textwidth]{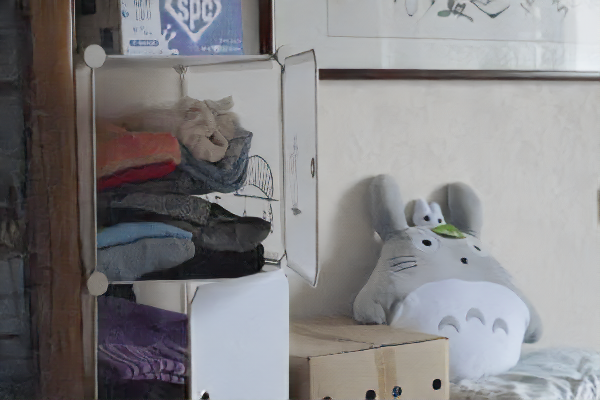}~&
			\includegraphics[width=.23\textwidth]{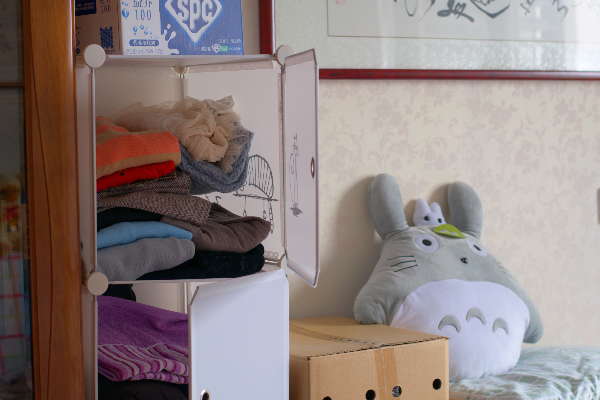}\\
			\includegraphics[width=.23\textwidth]{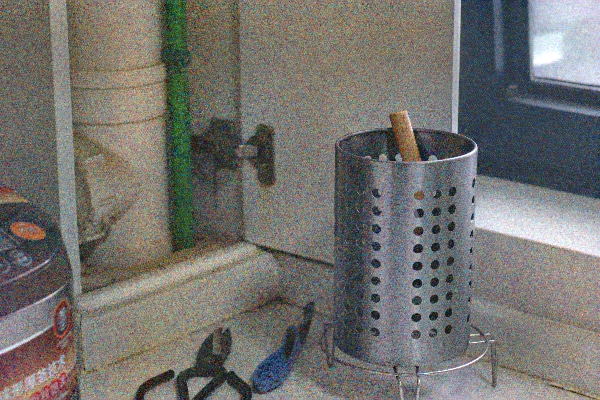}~&
			\includegraphics[width=.23\textwidth]{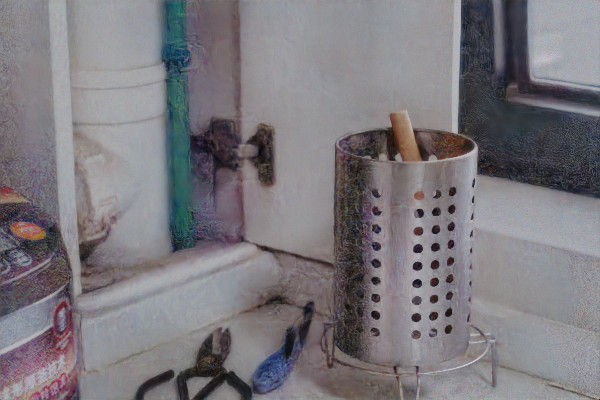}~&\includegraphics[width=.23\textwidth]{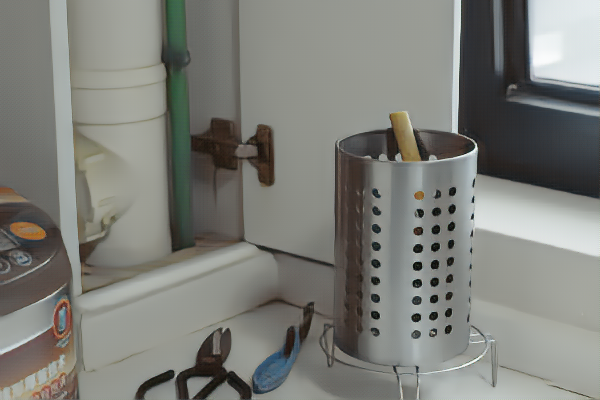}~&
			\includegraphics[width=.23\textwidth]{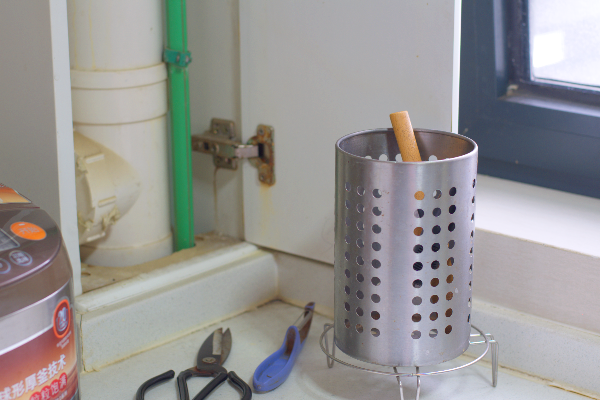}\\
			(a) EnlightenGAN~\cite{jiang2021enlightengan}~& (b) KinD~\cite{zhang2019kindling}~&(c) Ours~& (d) GT\\
		\end{tabular}
	\end{center}
	\caption{Examples of enhancement results on the LOL dataset synthesized with Poisson noise $\lambda=10$ (first row) and Gaussian noise $\sigma = 5$ (second row). The input low-light image (a), the estimated results of EnlightenGAN~\protect\cite{jiang2021enlightengan} (b), KinD~\protect\cite{zhang2019kindling} (c), Ours training with original LOL image pairs (d) the ground truth (e), respectively. Please \textbf{ZOOM IN} to see the details.}
	\label{fig:poisson10}
\end{figure*}

\begin{figure*}[!t]
\centering
\includegraphics[width=1.\linewidth]{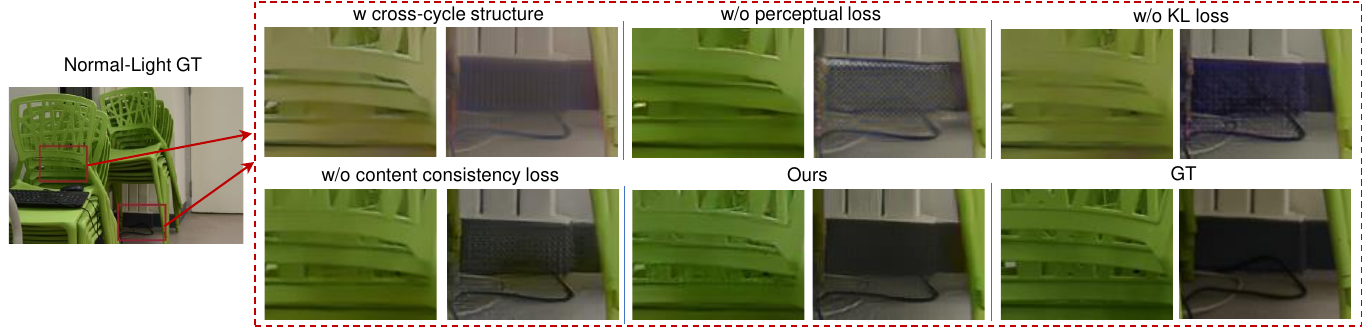} 
\caption{ One example of the result of with external cross-cycle structure,  the result without the perceptual loss, the result without the KL loss, the result without the content consistency loss, and the result of our complete model. (left is the corresponding normal-light ground truth)} 
\label{fig:ablation} 
\end{figure*}
\begin{table}[!t]
\centering
\footnotesize
\setlength{\tabcolsep}{0.5em}
\caption{Quantitative results on LOL dataset. -M denotes training with misaligned image pairs. }
\adjustbox{width=0.8\linewidth}{
    \begin{tabular}{l|ccc}
        \toprule
         Method & PSNR $\uparrow$  & SSIM $\uparrow$ & LPIPS $\downarrow$\\\midrule
           LIME~\cite{guo2016lime} &  14.02 & 0.56 & 0.35 \\
    Retinex-Net\cite{Chen2018Retinex}  & 16.77 &  0.56 & 0.35\\
        EnlightenGAN~\cite{jiang2021enlightengan} & 17.48 & 0.65 & 0.32\\
     Zero-DCE~\cite{guo2020zero} & 14.86 & 0.54 &  0.34\\
        LR3M~\cite{ren2020lr3m} & 18.91 & 0.75 & 0.28 \\
       RUAS~\cite{liu2021ruas} & 18.23& 0.72 &0.35 \\
    KinD~\cite{zhang2019kindling} & 20.38 &  0.80 & 0.17\\
    DRBN~\cite{Yang_2020_CVPR} & 20.08 & 0.83 & 0.16 \\
        CSDNet~\cite{ma2021learning} & 21.63 & \textcolor{red}{\textbf{0.85}} & 0.19\\
    Zhao~\etal~\cite{Zhao_2021_ICCV} & 21.71 & 0.83 & 0.20\\
    Lv~\etal~\cite{lv2021attention} & 20.24& 0.79 & \textcolor{blue}{0.14}  \\
     MIRNet~\cite{Zamir2020MIRNet} & \textcolor{blue}{24.14} & \textcolor{blue}{0.84} & \textcolor{red}{\textbf{0.13}} \\
     \rowcolor{Gray} Ours & \textcolor{red}{\textbf{25.25}} & \textcolor{red}{\textbf{0.85}} & \textcolor{red}{\textbf{0.13}}\\\midrule
            Retinex-Net~\cite{Chen2018Retinex}-M& 14.98 & 0.43 & 0.47  \\
       KinD~\cite{zhang2019kindling}-M& \textcolor{blue}{18.52} & \textcolor{blue}{0.62} & \textcolor{blue}{0.40}\\
           DRBN~\cite{Yang_2020_CVPR}-M & 17.58 & 0.59 & 0.49 \\
     \rowcolor{Gray} Ours-M & \textcolor{red}{\textbf{25.22}} & \textcolor{red}{\textbf{0.83}} & \textcolor{red}{\textbf{0.14}}\\
        \bottomrule
    \end{tabular}
}
\label{tab:lol_res}
\end{table}

\vspace{1mm}
\noindent\textbf{Other public aligned datasets.} 
We then show that our method can be extended to aligned dataset like LOL~\cite{Chen2018Retinex}, a dataset with 500 perfectly aligned $400\times600$ low/normal-light image pairs. Following the original partitions, we employ 485 image pairs as the training set and 15 image pairs as the testing set. \renjie{We first work with the original LOL dataset with aligned pixels. Then, we randomly simulated up to 10 pixels offsets for the training image pairs, to evaluate the robustness of low-light enhancement methods subjective to pixel shifts in a more controlled setting (The corresponding results using each method is appended with ``-M" in Table~\ref{tab:lol_res}). For image pairs in the evaluation dataset, there is always a corresponding aligned normal-light ground truth for the low-light images.} Table~\ref{tab:lol_res} lists the quantitative results among the competitors on LOL dataset\footnote{\adds{We adjust the illumination of the enhanced results over LOL dataset according to the truly average illumination for better measurement following the recent works~\cite{zhang2019kindling}, \cite{zhang2021beyond} and \cite{wang2021low}.}}. Our method outperforms all the other methods in either case perfectly aligned setting or misaligned setting. 
Figure~\ref{fig:cleanlol} compares some examples of the enhanced images by the proposed CIDN to those by the competing methods. With or without simulated pixel shifts, the proposed CIDN can always enhance images to higher visual quality.



\subsection{Network Analysis}
\vspace{1mm}
\noindent\textbf{Noise robustness.}
To further evaluate the robustness \renjie{of our CIDN to different noise quantitatively, we simulate Gaussian noise with $\sigma=5, 10$ and Poisson noise with $\lambda=10, 15$ into the low-light images of LOL~\cite{Chen2018Retinex} evaluation set, respectively.} We choose one unsupervised method EnlightenGAN~\cite{jiang2021enlightengan}, and one fully supervised method KinD~\cite{zhang2019kindling} as competitors.
No pixel shift is simulated, and the training data are noise-free.
Table~\ref{tab:noisy_res} reports the quantitative enhancement results using our CIDN and other competing methods on the noisy LOL dataset. 
Our method consistently provides the best results for all noise types and levels.
Figure~\ref{fig:poisson10} demonstrates the enhancement results with synthesized Possion noise and more visual examples of the enhanced images are included in the supplementary document.
The superior noise-robustness of CIDN is due to the effective modeling of the cross-image brightness and the image content. Noise is independent of neither image nor brightness feature, thus would not survive in our CIDN modeling.

\begin{table}[!t]
    \centering
    \footnotesize
        \caption{Quantitative results on synthesized noisy LOL evaluation set. The best result is in
    \textbf{bold}.}
 
    \setlength{\tabcolsep}{0.3em}
    \adjustbox{width=\linewidth}{
    \begin{tabular}{l|ccc|ccc}
    \toprule
    \multirow{2}{*}{Method} &  \multicolumn{3}{c}{$\sigma =5$} &  \multicolumn{3}{c}{$\sigma =10$}  \\
         & PSNR$\uparrow$  & SSIM $\uparrow$ & LPIPS$\downarrow$ &  PSNR$\uparrow$  & SSIM$\uparrow$  & LPIPS$\downarrow$\\\midrule
    EnlightenGAN~\cite{jiang2021enlightengan} & 15.31 & 0.41 & 0.56 & 14.80 & 0.27 & 1.01  \\
    KinD~\cite{zhang2019kindling} & 19.81 & 0.70 & 0.38 & 17.45 & 0.39 & 0.74 \\
    \rowcolor{Gray} Ours & \textbf{22.02} & \textbf{0.79} & \textbf{0.27} & \textbf{19.88} & \textbf{0.63} & \textbf{0.48}\\
    \midrule
      \multirow{2}{*}{Method} &  \multicolumn{3}{c}{$\lambda =10$} &  \multicolumn{3}{c}{$\lambda =15$} \\
         & PSNR$\uparrow$  & SSIM$\uparrow$  & LPIPS$\downarrow$  &  PSNR$\uparrow$  & SSIM$\uparrow$  & LPIPS$\downarrow$\\\midrule
    EnlightenGAN~\cite{jiang2021enlightengan} & 16.39 & 0.56 & 0.47 & 15.81 & 0.52 & 0.57 \\
    KinD~\cite{zhang2019kindling}  &  18.98 & 0.80 & 0.29 & 18.39 & 0.78 & 0.33\\
    \rowcolor{Gray} Ours  & \textbf{22.43} & \textbf{0.86} & \textbf{0.20} & \textbf{21.79} & \textbf{0.83} & \textbf{0.21}\\
    \bottomrule
    \end{tabular}}

    \label{tab:noisy_res}
\end{table}

\begin{table}[!t]
    \centering
    \footnotesize
    \caption{Quantitative evaluation results on NM2L dataset for the model without the content consistency loss, the model without the perceptual loss, the model without the KL loss, the model with external cross-cycle structure, and the complete model. The best result is in \textbf{bold}.}
    \setlength{\tabcolsep}{0.4em}
    \adjustbox{width=0.9\linewidth}{
    \begin{tabular}{c|ccc}
    \toprule
     & PSNR $\uparrow$ & SSIM $\uparrow$  &LPIPS $\downarrow$ \\
    \midrule
        w/o content consistency loss & 21.13 & 0.79 & 0.17\\
     w/o perceptual Loss & 19.36 & 0.75 & 0.23\\
      w/o KL loss& 19.98 & 0.79& 0.21  \\
     w cross-cycle structure & 17.98 & 0.75 & 0.24\\
         \adds{ w/o adversarial loss} & \adds{ 17.08} & \adds{ 0.71} & \adds{ 0.33}\\
    \rowcolor{Gray} Complete model & \textbf{21.58} &\textbf{0.81} & \textbf{0.16}\\
    \bottomrule
    \end{tabular}
    }
    \label{tab:ablation}
\end{table}

		\begin{figure}[t]
	\begin{center}
		\begin{tabular}{c@{ }c@{ }c}
			\includegraphics[width=.3\linewidth,height=.22\linewidth]{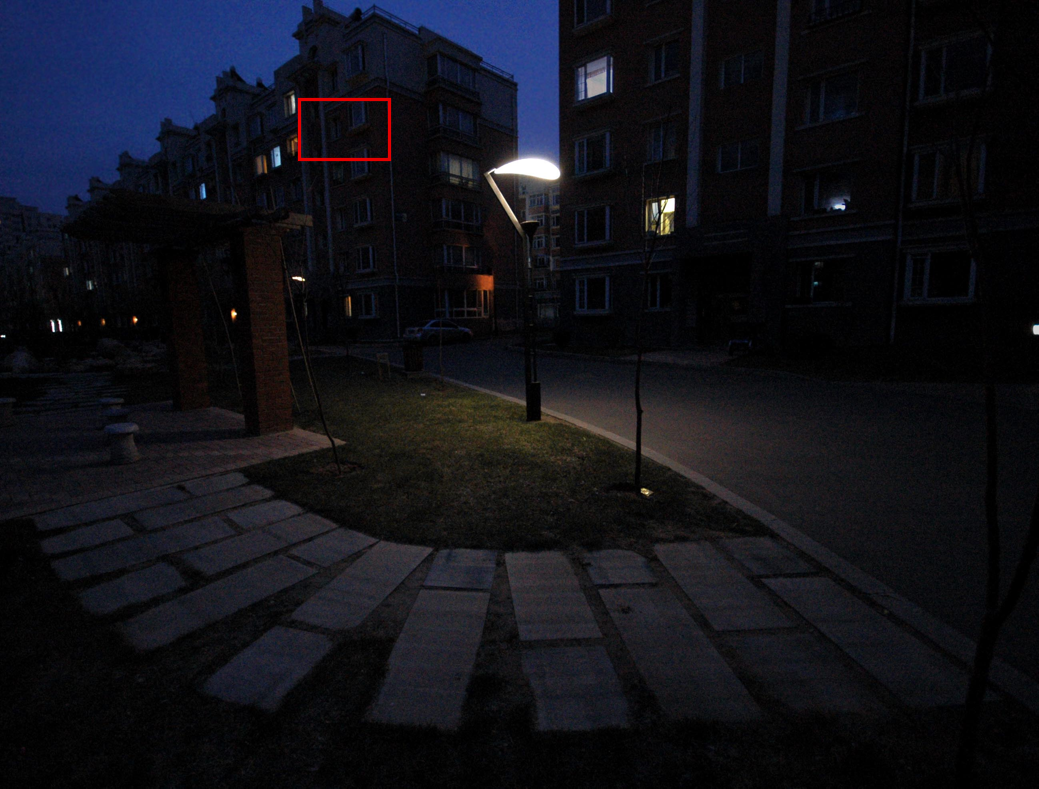}~&
			\includegraphics[width=.3\linewidth,height=.22\linewidth]{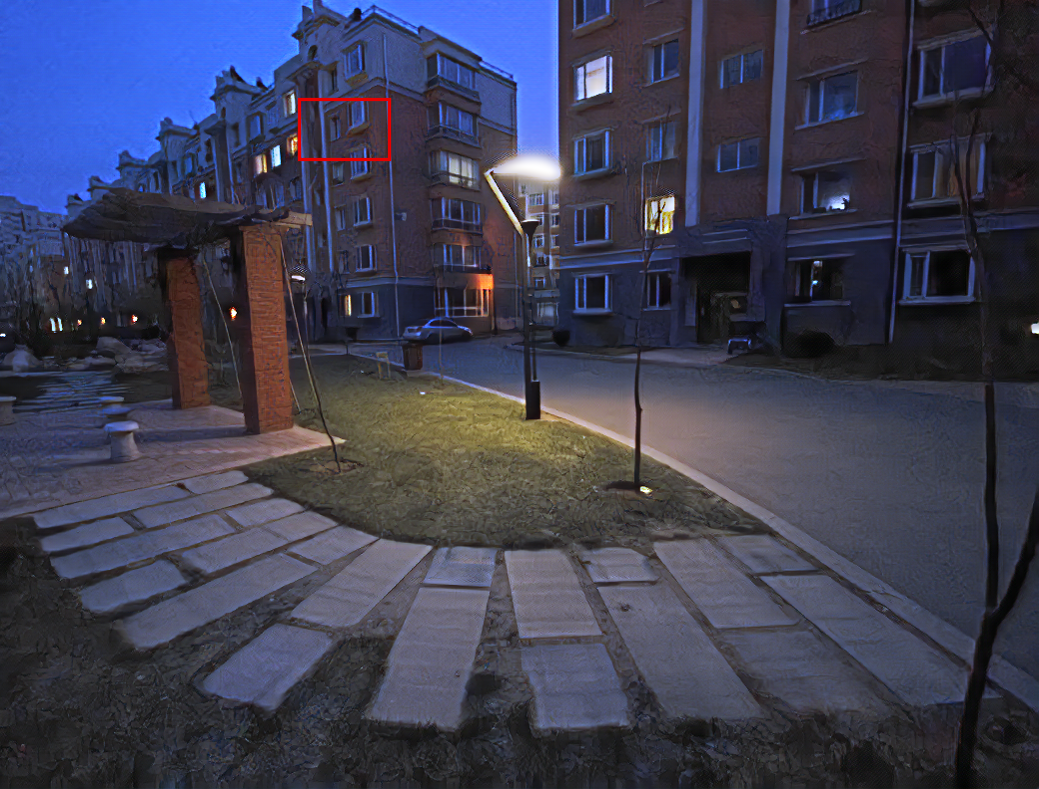}~&
			\includegraphics[width=.3\linewidth,height=.22\linewidth]{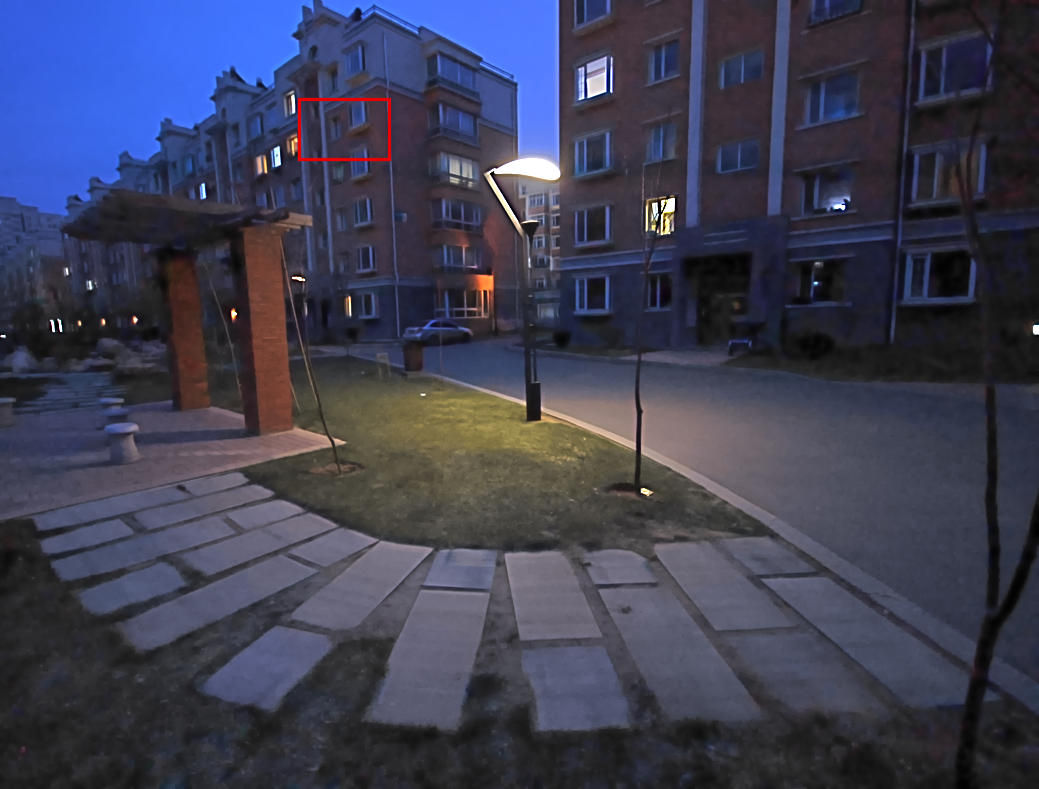}\\
				\includegraphics[width=.3\linewidth,height=.22\linewidth]{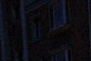}~&
			\includegraphics[width=.3\linewidth,height=.22\linewidth]{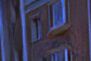}~&
			\includegraphics[width=.3\linewidth,height=.22\linewidth]{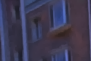}\\
			Input & LOL & NM2L\\
		\end{tabular}
	\end{center}
	\caption{Examples of input low-light images and enhancement results by CIDN training on LOL and NM2L datasets (first row), and their corresponding zoom-in regions (second row).}
	\label{fig:dataset_power}
\end{figure}

		\begin{figure}[t]
	\begin{center}
		\begin{tabular}{c@{ }c}
			\includegraphics[width=.45\linewidth]{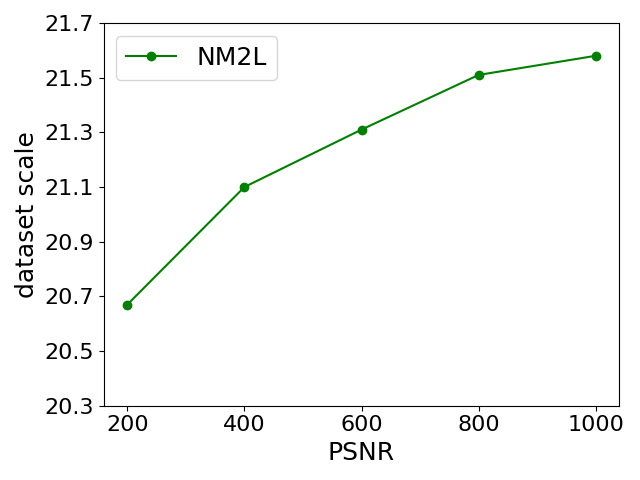}~&
			\includegraphics[width=.45\linewidth]{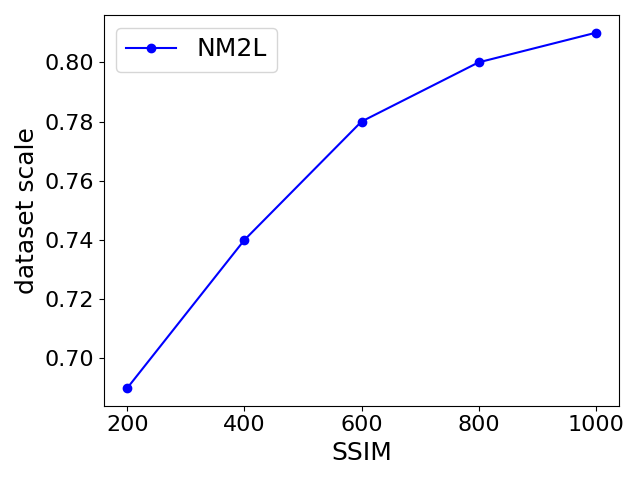}\\
		\end{tabular}
	\end{center}
	\vspace{-0.5cm}
	\caption{The evaluation results on the testing set of NM2L in PSNR (left) and SSIM (right) as training data increases. }
	\label{fig:dataset_power2}
\end{figure}

\vspace{1mm}
\noindent\textbf{Ablation study.} 
Furthermore, we conduct several experiments to evaluate the effectiveness of different training losses used for CIDN. We first remove the content consistency loss. From the results shown in Figure~\ref{fig:ablation}, the final estimated images without the content consistency loss cannot effectively suppress the noise and artifacts.
We then remove the perceptual loss.
As shown in Figure~\ref{fig:ablation}, \renjie{the structural details cannot be well preserved in the results obtained without the perceptual loss.}
We also remove the KL loss. 
The structure information still cannot be preserved entirely without the KL loss, which means the extracted brightness features are not independent of structure information, as shown in Figure~\ref{fig:ablation}.
\adds{The adversarial loss constrains the extracted brightness feature is illumination-aware. Thus the results without adversarial loss would destroy the underlying illumination of output images.}
\adds{Besides, inspired by the success of cross-cycle consistency constraint~\cite{lee2018diverse} in image translation, we also exploit the disentangled brightness and content features for cyclic reconstruction instead of one-path reconstruction.
	Specifically, for the enhanced result $\tilde{\mathbf{y}}$ and the darkened result $\tilde{\mathbf{x}}$ from normal-light and low-light decoders, we apply the second disentanglement to extract the brightness and content features. After second brightness feature swapping, we achieve the reconstructed ${\tilde{\mathbf{x}}}^{\prime}$ and ${\tilde{\mathbf{y}}}^{\prime}$ as following:
	\begin{equation}
    {\tilde{\mathbf{y}}}^{\prime} = G_Y (E_C(\tilde{\mathbf{x}}) , E_B(\tilde{\mathbf{y}}))\;\quad
      {\tilde{\mathbf{x}}}^{\prime} = G_X (E_C(\tilde{\mathbf{y}}) , E_B(\tilde{\mathbf{x}}))\;.
\end{equation}
We then employ $\ell_1$ loss to minimize the error map between \{$\mathbf{x}$, $\mathbf{y}$\} and \{${\tilde{\mathbf{x}}}^{\prime}$, ${\tilde{\mathbf{y}}}^{\prime}$\}.}
From the results shown in Figure~\ref{fig:ablation}, an external cross-cycle structure may lead to color distortions.
The quantitative values in Table~\ref{tab:ablation} also prove the effectiveness of our framework.

\begin{figure}[!t]
	\begin{center}
		\begin{tabular}{c@{ }c}
			\includegraphics[width=.48\linewidth]{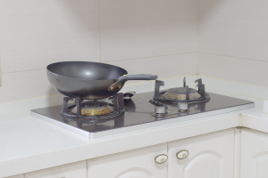}~&
			\includegraphics[width=.48\linewidth]{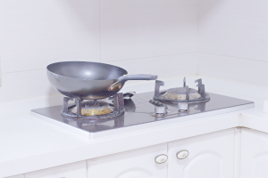}\\
				(a) Over-exposure input~& (b) Zero-DCE~\cite{guo2020zero}\\
			\includegraphics[width=.48\linewidth]{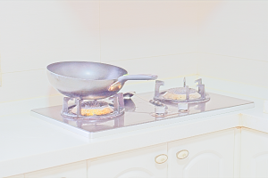}~&
			\includegraphics[width=.48\linewidth]{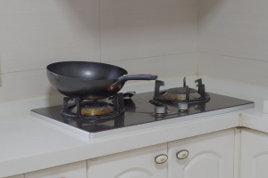}\\
					(c) EnlightenGAN~\cite{jiang2021enlightengan}~& (d) Photoshop\\
			\includegraphics[width=.48\linewidth]{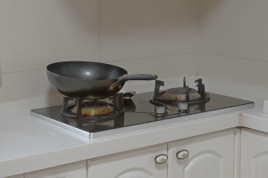}~&
			\includegraphics[width=.48\linewidth]{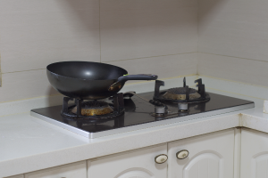}\\
					(e) Ours~& (f) Properly exposed ref.\\
		\end{tabular}
	\end{center}
\caption{ One example of over-exposure correction. (a) over-exposure input, (b) the result of Zero-DCE~\cite{guo2020zero}, (c) the result of EnlightenGAN~\cite{jiang2021enlightengan}, (d) the result processed by Photoshop, (e) the result of proposed model, and (f) properly exposed reference.} 
\label{fig:overexposure} 
\end{figure}

\vspace{1mm}
\noindent\textbf{The power of large-scale dataset.} 
\adds{
To demonstrate the power of the proposed dataset NM2L, we evaluate the proposed CIDN model with different data scale of training set.
Since the proposed NM2L and the public LOL datasets have different distribution taken by different cameras and under different conditions, it is unfair to evaluate the effect of the data scale of dataset on the testing set of NM2L or LOL.
	To this end, we employ the non-reference dataset LIME and MEF as the benchmark.
	Figure~\ref{fig:dataset_power} illustrates the visual examples of the enhanced results of the proposed CIDN training on LOL, NM2L, and LOL+NM2L datasets.
	The enhanced result for some low-quality input from the model training on NM2L would be better than training on LOL since the NM2L considered the extra noise when collecting low-light images.
	The combination of LOL and NM2L datasets would boost the enhancement performance since the various samples with diverse scenes and brightness conditions in training stage would boost the performance when transferring to the unseen testing set.
	Moreover, we also provide the quantitative results to further verify the large-scale training set might boost the enhancement performance.
	As shown in Figure~\ref{fig:dataset_power2}, the enhanced results would be better if employing more training pairs, which means the 1000 image pairs are more powerful than 500 image pairs for enhancement model training.
}

\subsection{Exposure Correction} 
\add{Capturing photographs with \renjiess{inappropriate} exposures remains a major source of errors in camera-based imaging.
Exposure problems are categorized as either: (1) under-exposure, where the exposure is too short, resulting in dark regions, or (2) over-exposure, where the camera exposure is too long, resulting in bright and washed-out image regions.
Both our focused low-light image restoration and under-exposure image correction are adjusting images with low brightness to normal brightness.
We also run our model on some over-exposure images.
Note that in this experiment, we directly apply the pretrained model training on NM2L dataset, which only contains low/normal-light image pairs, without further tuning or re-training on any exposure correction dataset.
Our method can be directly applied to the exposure correction task without any finetuning or re-training, while competing methods fail, as shown in Figure~\ref{fig:overexposure}. }

\begin{figure}[!t]
	\begin{center}
		\begin{tabular}{c@{ }c}
			\includegraphics[width=.48\linewidth]{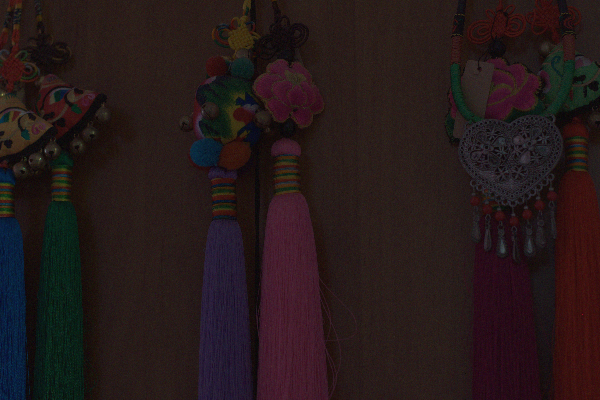}~&
			\includegraphics[width=.48\linewidth]{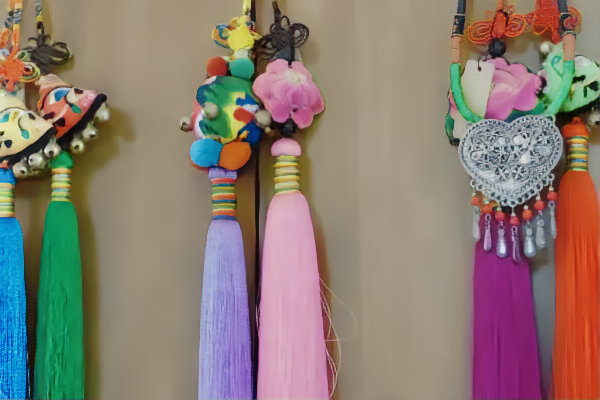}\\
				(a) Input~& (b) Our result\\
			\includegraphics[width=.48\linewidth]{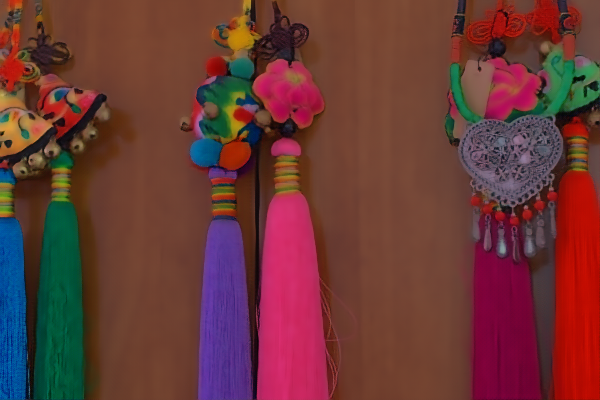}~&
			\includegraphics[width=.48\linewidth]{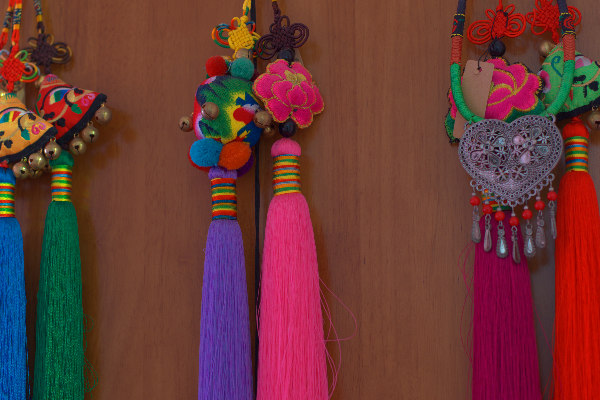}\\
					(c) Our adjusted result~& (d) GT\\
		\end{tabular}
	\end{center}
\caption{ One example of color cast case. (a) input low-light image with color cast, (b) the result of our proposed model, (c) the adjusted result by adjusting the intensity rate between RGB channels, and (d) corresponding ground truth.} 
\label{fig:limitation} 
\end{figure}

\subsection{Limitation Analysis}
\add{
\renjiess{Though our method achieves promising results in challenging cases, it may still lead to the color casting problem in some specified situations,} \eg, low-light images taken under red light conditions.
As shown in Figure~\ref{fig:limitation}(b), our network cannot \renjiess{recover} the color inside the cabinet region \renjiess{since} partial color information has been disentangled with brightness.
One way to alleviate this problem is selecting a  normal-light image \renjiess{with a similar color style} as guidance.
Alternatively, the color cast \renjiess{problems} can also be addressed by post-processing. Specifically, although the original low-light input is taken under the \renjiess{weak} illumination condition, the color information is still preserved.
Thus, the color cast can be corrected by adjusting the intensity rate between RGB channels based on the intensity rate of original low-light images, as shown in Figure~\ref{fig:limitation}(c).
}

\section{Conclusion}
In this work, we propose a method to solve the low-light image enhancement problem, which aims to correct the brightness and suppress image artifacts for better visibility.
By employing misaligned image pairs as the training data, we propose a Cross-Image Disentanglement Network to decouple cross-image brightness and content features. It then simultaneously corrects the brightness and suppresses artifacts in the feature domain based on the encoded low-dimensional features.
Our experimental results show promising results for our misaligned dataset and other popular low-light datasets.
Furthermore, we demonstrate that our method has superior noise suppression performance on both real and synthesized noisy data.

\ifCLASSOPTIONcaptionsoff
  \newpage
\fi



\bibliographystyle{IEEEtran}
\bibliography{myref}

\end{document}